%% file: main.tex
\documentclass[%
superscriptaddress,
preprint,
showpacs,preprintnumbers,
amsmath,amssymb,
aps,
pra,
longbibliography]{revtex4-1}

\usepackage{graphicx}
\usepackage{dcolumn}
\usepackage{bm}
\usepackage{mathtools}
\usepackage{extarrows} 
\usepackage{hyperref}
\usepackage{comment}
\usepackage{color}

\usepackage{array}
\newcolumntype{C}[1]{>{\centering\arraybackslash}m{#1}}
\newcolumntype{N}{@{}m{0pt}@{}}

\begin{document}


\title{The thermo-optic nonlinearity of single metal nanoparticles under intense continuous wave illumination}

\author{Ieng Wai Un}
\email{iengwai@post.bgu.ac.il}
\affiliation{School of Electrical and Computer Engineering, Ben-Gurion University of the Negev, Israel.}
\author{Yonatan Sivan}%

\affiliation{School of Electrical and Computer Engineering, Ben-Gurion University of the Negev, Israel.}

\date{\today}

\begin{abstract}
Over the last few decades, extensive previous studies of the nonlinear response of metal nanoparticles report a wide variation of nonlinear coefficients, thus, revealing a highly confused picture of the underlying physics. This naturally prevents rational design of these systems for practical devices. Here, we provide a systematic study of the nonlinear response of metal spheres under continuous wave illumination within a purely thermal model, i.e., whereby the illumination only acts to modify the optical and thermal parameters via their dependence on the temperature. We characterize the strong dependence of the temperature rise and overall thermo-optic nonlinear response on the particle size and permittivity, on the optical and thermal host properties, as well as on the thermo-derivatives of these properties. This dependence on the non-intrinsic parameters explains why it is inappropriate to extract an intrinsic nonlinear coefficient from a specific system; equivalently, it explains the large differences in reported values for such systems, as well as for more complicated metal-dielectric systems and even pulsed illumination schemes. Despite the revealed complex multi-parameter dependence, we managed to uncover a rather simple behaviour of the nonlinear response. In particular, we show that the nonlinearity coefficients exhibit a dependence on the illumination intensity which mimics the dependence of the temperature itself on the illumination intensity, namely, it grows for small nanoparticle sizes, reaches a maximum and then decreases monotonically for larger nanoparticles. The improved modelling allows us to demonstrate an overall nonlinear response which is about a 1000 times higher than in other strongly nonlinear systems (e.g., $\epsilon$-near-zero systems); it also provides an excellent match to experimental measurements of the scattering from a single metal nanoparticles, thus, confirming the dominance of the thermal nonlinear mechanism. Our work lays the foundations for an overall evaluation of previous studies of the nonlinear response of metal-dielectric system under general conditions.
\end{abstract}

\pacs{42.25.Bs, 42.25.Fx, 65.80.-g, 66.70.Df, 78.20.N-, 78.20.Nv}

\maketitle


\section{Introduction\label{sec:intro}}

Metals are well-known for their ability to absorb light efficiently and for the consequent heating~\cite{thermo-plasmonics-review}. Understanding how the generated heat affects the thermal and optical properties of metals to intense illumination is, thus, of fundamental importance, and is also significant for a wide range of applications, especially in biology and energy harvesting, e.g., photo-thermal imaging~\cite{Shaked-PT_imaging,PT_imaging,Cognet-phtthm-meth-AnalyChem}, photothermal therapy~\cite{Riley-AuNP-PTT-review-2017,MBCortie-AuNP-hyperthermia-2018,Vines-AuNP-PTT-review-2019}, thermo-photovoltaics~\cite{Zubin_thermophotovoltaics,thermal_emission_vlad}, plasmonic-heating-induced nanofabrication~\cite{Hashimoto-plasmon-heat-fab-glass-2016,Hashimoto-plasmon-nanofab-2016}, water boiling and bubble generation~\cite{Halas-bubble1,Halas-bubble2,Baffou-bubble1,Orrit-bubble,Lukianova-bubble}, nanoscale phase transition~\cite{Urban-nm-phase-trans,Bendix-NPtemp-measure} and plasmon-assisted photocatalysis~\cite{Baffou_solvothermal,Dubi-Sivan-Faraday,anti-Halas-Science-paper,Y2-eppur-si-riscalda,thm_hot_e_faraday_discuss_2019,dyn_hot_e_faraday_discuss_2019,Liu-Everitt-Nano-Letters-2019}.

The thermo-optic response to illumination depends on two elements. First, on the dependence of the metal permittivity on the temperature, clearly an inherent material property. Second, on the dependence of the temperature rise itself on non-intrinsic parameters such as the illumination parameters, the particle geometry, and the optical and thermal properties of its surroundings; this dependence becomes non-trivial for sufficiently high illumination intensities and large nanoparticles (NPs)~\cite{thermo-plasmonics-basics,thermo-plasmonics-review,Un-Sivan-size-thermal-effect}. The heating can be particularly efficient if the illumination is tuned to a resonance of the metal nanoparticle 
where the strong field confinement gives rise to strong absorption.

Early studies of this problem were concerned with ensembles of NPs under illumination at various different durations (see e.g.,~\cite{Blau_review_SA_RSA} for a review). Later, the attention was directed to the study of single NPs~\cite{Orrit_single_NP_review}. The vast majority of previous studies focused on the thermo-optic nonlinear response of metal nanostructures under {\em ultrafast} illumination~\cite{Del_Fatti_UF_elec_dyn_opt_NLTY_MNP_JPCB_2001,Del-Fatti-PRL-UF-NLTY-MNR-2011,Stoll_review}. In this case, the temperature rise due to absorption of light is inversely proportional to the heat capacity of the electrons, and is determined by the interplay between absorption of incident photons by electrons, thermalization due to electron-electron interactions and energy exchange with the phonons. Standard experimental signatures include differential reflectivity, transmissivity, scattering etc. or even transient frequency changes due to permittivity changes. In the first picosecond or so, the nonlinear response is dominated by the electron dynamics; accordingly, it is typically weak. At later stages, the stronger sensitivity of the permittivity to the (indeed smaller) rise of the phonon temperature becomes dominant. The overall nonlinear response is proportional to the particle volume and is non-local (delayed) in time~\cite{Stoll_review,Biancalana_NJP_2012} and in space~\cite{ICFO_Sivan_metal_diffusion,Sivan_Spector_metal_diffusion}; the latter effect (namely, heat diffusion) is very strong within the metal structure, causing the temperature of the nanostructure to become uniform on a subpicosecond time scale~\cite{Baffou_pulsed_heat_eq_with_Kapitza,thermo-plasmonics-review,Sivan_Spector_metal_diffusion}. In contrast, the heat transfer from the metal to the surrounding is slower, such that it was usually neglected in ultrafast studies. Comprehensive theoretical and experimental description of the nonlinear thermo-optic response in this regime is provided in~\cite{optic_excitation,delFatti_nonequilib_2000,Del_Fatti_UF_elec_dyn_opt_NLTY_MNP_JPCB_2001,Del_Fatti_UF_ee_sca_eng_exchg_2004,Stoll_review,Aeschliman_e_photoemission_review,Langbein_measure_plasmon_dyn_AuNP_PRB_2012}. In this ultrafast regime, there is also an instantaneous coherent nonlinear response that is not usually associated with heat, but rather with non-thermal electrons. It, in general, leads to frequency conversion~\cite{Orrit-THG-sng-gold-NP-NanoLett,Zayats-NL-plasmonics-review-NatPhoton,Italians_double_resonance,Taiwanese_double_resonance_films}; these (generally weaker) effects will not be studied in the current paper.

For timescales of 10 picoseconds to several hundreds picoseconds, the electronic response is of lesser importance, and the heat dissipation from the NP to its environment cannot be neglected~\cite{Del_Fatti_cool_dyn_thm_intf_resis_PRB_2009,Stoll_environment,Matthew_acoustic_optic_metal_NP_acsnano_2017,Gandolfi_UF_Thm_opt_MNP_JPCC_2018}. The cooling dynamics also causes the NP to expand, such that rapid acoustic oscillations are induced~\cite{Crut_acoustics,Haim_acoustics,PT_PA_imaging_Danielli,Pelton_acoustics} and complex bubble formation may ensue~\cite{Baffou-bubble1,Orrit-bubble}. The analysis of these effects will also not be studied in the current paper.

In contrast, under CW illumination, the temperature dynamics in metals is independent of the heat capacity, and instead, is determined by the interplay between absorption of incident photons and heat diffusion away from the NP~\cite{thermo-plasmonics-review}. In that sense, the temporal non-locality is irrelevant but the spatial non-locality is dominant. One of the consequences of this is that the dependence of the nonlinear response on the NP size differs from that under ultrafast illumination - it scales as the surface area for small NPs, and then exhibits a complex oscillatory behaviour for larger NPs, see~\cite{Un-Sivan-size-thermal-effect}. In this regime, non-thermal effects (hence, coherent nonlinearities) are negligible compared with effects associated with thermalized electrons~\cite{Dubi-Sivan,Dubi-Sivan-Faraday}; this justifies the use of the permittivity data under external heating. 

The plethora of studies discussed so far provide a wide range of very different values for the nonlinear response, see for example~\cite{Boyd-metal-nlty}. This reveals a highly confused picture of the underlying physics which naturally prevents rational design of these systems for practical applications. At least partially, the variation in reported nonlinear coefficients has its root in the large differences of nanoparticle sizes and shapes, as well as in the illumination wavelength, intensity and duration and host properties. It is the goal of this manuscript to explain this variability in results, and to provide a simpler systematic way to the characterize the nonlinear response of metal NPs.

Recently, it has been shown experimentally that single metal NPs exhibit strong and non-trivial nonlinear scattering and absorption under CW illumination~\cite{plasmonic-SAX-PRL,plasmonic-SAX-ACS_phot,plasmonic-SAX-rods-Ag,SUSI,Jagadale_SWChu_NL_abs_sca_2019,Hashimoto-nanoscale-cooling,Hashimoto_Opt_Sca_Spec_Thermometry}. Specifically, the authors demonstrated that the normalized scattering and absorption decrease when the illumination intensity increases. These effects were referred to as saturation of scattering/absorption~\cite{plasmonic-SAX-PRL,plasmonic-SAX-ACS_phot,plasmonic-SAX-rods-Ag,SUSI,Jagadale_SWChu_NL_abs_sca_2019}. It has been also demonstrated that the amount by which the scattering/absorption decrease strongly depends on the substrate medium and on the contact geometry between the NP and the substrate~\cite{Hashimoto-nanoscale-cooling,Hashimoto_Opt_Sca_Spec_Thermometry}. The intensity-dependence of the scattering has demonstrated potential for applications in super-resolution imaging~\cite{plasmonic-SAX-PRL,plasmonic-SAX-OE,SUSI} and all-optical switching~\cite{SUSI}. When the illumination intensity is sufficiently high, the decrease of the scattering/absorption changes to an increase, an effect referred to as ``reverse saturation of scattering''. 

Following these experimental works, we have embarked upon a study of the thermo-optic nonlinearity based on the thermo-optic effect. In~\cite{Sivan-Chu-high-T-nl-plasmonics}, we demonstrated a qualitative agreement between the experimental data and a numerical calculation performed under the quasi-static approximation (namely, for uniform field and subwavelength NPs). In~\cite{Gurwich-Sivan-CW-nlty-metal_NP}, we complemented the numerical simulations of~\cite{Sivan-Chu-high-T-nl-plasmonics} with some analytic insights and characterized various physical configurations. Thes studies showed that few nm metal spheres exhibit extremely large nonlinear thermo-optic response under continuous wave illumination which beyond a $\sim 100 $K rise in temperature has to be described by a non-perturbative model. We studied the interplay between the optical parameters of the metal (e.g., the resonance quality), its geometry and the optical and thermal properties of the host. 

Here, encouraged by the success of~\cite{Sivan-Chu-high-T-nl-plasmonics}, we go beyond the quasi-static approximation, by modelling the thermo-optic response for larger NPs without any approximation of the electromagnetic response. We calculate the temperature, permittivity, local-field and scattering cross-section of the NPs by using the best available experimentally measured data of the temperature-dependent permittivity~\cite{PT_Shen_ellipsometry_gold,Shalaev_Ag_permittivity_ACSPhotonics}. Then, we characterize the thermo-optical nonlinearity of a single metallic NP, namely, by combining the dependence of the temperature and scattering response on the illumination intensity with the thermo-derivatives of the various optical and thermal properties. This analysis points to reasons for the variety of reported values of the nonlinear optical response of metal NPs and shows that ultrafast analysis should be handled with care when used to explain the CW response.

This study is not only an important step towards verifying the hypothesis~\cite{Sivan-Chu-high-T-nl-plasmonics,Gurwich-Sivan-CW-nlty-metal_NP} that the thermo-optical nonlinearity is responsible for the strong nonlinear scattering observed in~\cite{plasmonic-SAX-PRL,plasmonic-SAX-ACS_phot,plasmonic-SAX-rods-Ag,SUSI} for NPs of large size, but also an indispensable step towards showing the importance of the thermo-optical nonlinearity in plasmon-assisted photocatalysis~\cite{dyn_hot_e_faraday_discuss_2019,thm_hot_e_faraday_discuss_2019,Y2-eppur-si-riscalda}. In particular, these results are a first step towards elucidating the errors in the claims in~\cite{Halas_Science_2018_response} about the origin of the nonlinear thermo-optic response in plasmon-assisted photocatalysis experiments, see a detailed discussion in~\cite{R2R}. Finally, our analysis is also relevant for other absorbing materials like graphene, semiconductors~\cite{Lewi_thm_opt_PbTe_meta_atom,Lewi_thm_reconfig_meta_atoms,Shi-Wei-Si-Nanoscopy,Shi-Wei-Si-NLTY} etc..

The paper is organized as follows. In Section~\ref{sec:config}, we first describe the configuration and the basic assumptions of the model; in Section~\ref{sec:nl_temp_sca}, we develop the model equations for the temperature within the NP; and in Section~\ref{sec:characterize_nlty}, we describe how to characterize the thermal-optic nonlinearity. We then proceed by several numerical examples (Section~\ref{sec:results}) and complement the numerical results with a detailed analysis in Section~\ref{sec:fur_analy}. This analysis elucidates the main result of this work, namely, that the nonlinear response reaches a maximal value for NP sizes of several tens of nms. It also identifies the role of each of the thermo-derivatives of the parameters in the heated NP system. Finally, we provide a discussion of the results and an outlook in Section~\ref{sec:discussion}.

\section{Configuration\label{sec:config}}
We consider a {\em single} spherical metal NP of radius $a$ with temperature-dependent permittivity $\varepsilon_m(T)$ in a loss-less dielectric host $\varepsilon_h$ illuminated by a high intensity CW plane wave. The absorption of incoming photons causes the NP to heat up, an effect which is balanced by heat transfer to the environment such that the temperature reaches a steady-state. In this case, the heat equation reduces to the Poisson equation,
\begin{align}
\nabla \cdot \left(\kappa({\bf r},T({\bf r})) \nabla T({\bf r}) \right) = - p_{\textrm{abs}}(\omega,{\bf r},T({\bf r})), \label{eq:poissoneq}
\end{align}
where $\kappa({\bf r},T({\bf r}))$ is the thermal conductivity (which is, in general, temperature-, hence, space-dependent) and $p_{\textrm{abs}}(\omega,{\bf r},T({\bf r}))$ is the absorbed power density. Here, we only consider one-photon absorption and neglect potential multi-photon absorption so that the absorbed power density is given by $p_{\textrm{abs}}(\omega,{\bf r},T({\bf r})) = \dfrac{\omega}{2} \varepsilon_0 \varepsilon^{\prime\prime}_m(\omega,T({\bf r})) |{\bf E}(\omega,{\bf r})|^2$, where ${\bf E}(\omega,{\bf r})$ is is the total (local) electric field~\footnote{Multi-photon absorption is unlikely for the moderately high intensities used for CW illumination. Its inclusion will anyhow cause only a modest quantitative change to the results of the current study.}. We also ignore the small differences between the electron and lattice temperatures for simplicity, see justification in~\cite{Abajo_nano-oven,Dubi-Sivan,Dubi-Sivan-Faraday}.

The illumination-induced heating of the NP causes a modification of the optical and thermal properties of the NP and its surrounding. This thermo-optic effect couples the equations for the temperature and electric fields, thus, giving rise to a nonlinear dependence of the absorption and scattering from the NP on the incident intensity. In the current work, we calculate the particle temperature and study the thermo-optic nonlinearity of a {\em single} NP as a function of its size, of the illumination wavelength and of the optical and thermal properties. 

Since we would like to concentrate on the interplay among temperature, particle size and thermo-optic nonlinearity, we avoid using sophisticated solid-state physics models of the temperature dependence of these properties (these are available in e.g.,~\cite{Stoll_review,Langbein_measure_plasmon_dyn_AuNP_PRB_2012,ICFO_Sivan_metal_diffusion}). Instead, we perform all the calculations below by utilizing the best available empirical data for the temperature-dependent permittivities extracted from recent ellipsometry measurements of single crystalline thin Au films (up to $600$ K in the wavelength range of $200 - 1680$ nm~\cite{PT_Shen_ellipsometry_gold}) and thin Ag films (up to $900$ K in the wavelength range of $330 - 2000$ nm~\cite{Shalaev_Ag_permittivity_ACSPhotonics}); similar data sets can be found in~\cite{Liljenvall_Ag_eps_JPC,Shalaev_ellipsometry_gold,Magnozzi_temp_dep_Ag_eps_PhysRevMaterials,Lazzari_Ag_diff_reflec_spect_Nanotech}. Specifically, labeling the ambient properties (i.e., the vanishing incident intensity limit) by a subscript 0, we model the dependence of the metal permittivity on the temperature by a second-order polynomial, namely,
\begin{align}\label{eq:epsm_2nd_poly}
\varepsilon_m(\omega,T) &= \varepsilon^{\prime}_{m,0}(\omega) + B_m^{\prime}(\omega)(T - T_{h,0}) + D^{\prime}_m(\omega)(T - T_{h,0})^2\nonumber\\
&+ i\left[ \varepsilon^{\prime\prime}_{m,0}(\omega) + B_m^{\prime\prime}(\omega)(T - T_{h,0}) + D_m^{\prime\prime}(\omega)(T - T_{h,0})^2 \right],
\end{align}
where $T_{h,0}$ is the ambient temperature, $B_m = B_m^{\prime} + i B_m^{\prime\prime}$ and $D_m = D_m^{\prime} + i D_m^{\prime\prime}$ are the first- and the second-order thermoderivatives of the permittivity, respectively. Comparing to the linear dependence of the permittivity on the temperature assumed in Ref.~\cite{Gurwich-Sivan-CW-nlty-metal_NP}, the second-order polynomial is able to fit the data to a higher level of accuracy. Similarly, we assume for the host thermal conductivity that $\kappa_h(T) = \kappa_{h,0} + B_{\kappa,h}(T-T_{h,0})$~\footnote{A second order thermoderivative of $\kappa_h$ is supposed to be included for the temperature rise of a few hundred degrees. However, as we shall see in Eq.~(\ref{eq:tempnpsc}) and Eq.~(\ref{eq:tempnp_approx}), the importance of the first-order thermoderivative of $\kappa_h$ to the nonlinear response is reduced by a factor 2. Accordingly, the importance of the second-order thermoderivative of $\kappa_h$ to the nonlinear response is reduced by a factor 3. This is because only a small region of the host medium is heated up by the nanoparticle. Therefore, for the simplicity, we neglect the second order thermoderivative of $\kappa_h$ here.} (see, e.g. data in Ref.~\cite{Hashimoto-nanoscale-cooling}). In the calculations below, we limit ourselves to a maximum temperature rise smaller than 400 K such that the detailed assumptions above hold, and so that sintering and melting of the metal, as well as damage or phase transitions in the host are avoided~\cite{japanese_size_reduction,Baffou-bubble1,Baffou-bubble2,Hashimoto-plasmon-nanofab-2016,Hashimoto-plasmon-heat-fab-glass-2016}. 

In order to generate a complete picture of the nonlinear thermo-optic response of metals, it is required to solve Eq.~(\ref{eq:poissoneq}) together with Maxwell's equations self-consistently since the electric field, the temperature, the absorbed power density and all temperature-dependent material parameters inside the NP are spatially non-uniform. However, while the electric field (hence, the  absorbed power density) non-uniformity might be significant (especially, as occurs for NPs of more than a few nm in size, for which high-order multipoles are excited), a comparison to exact simulations has shown that the non-uniformity of the temperature inside the NP is quite small (see~\cite{thermo-plasmonics-basics,Un-Sivan-size-thermal-effect}). The reason for that is that the thermal conductivity of the metal is typically much greater than the thermal conductivity of the host, $\kappa_m \gg \kappa_h$; we observe that this assumption is also valid in the presence of the thermo-optic nonlinearity. Furthermore, for the sake of simplicity, we neglect the temperature dependence of the host permittivity~\footnote{The temperature-dependence of $\varepsilon_h$ has the similar effect on the NP temperature as that of $\varepsilon_m^{\prime}$, i.e. shift the resonance wavelength. The relative importance of the thermoderivatives of $\varepsilon_h$ is substantially reduced since only a small region of the host medium is heated up by the NP, similar to the case of the host thermal conductivity.}. This will allow us to simplify the problem significantly and to obtain an approximate analytic solution of Eq.~(\ref{eq:poissoneq}). 

\section{Temperature of a single NP under CW illumination}\label{sec:nl_temp_sca}
The above assumptions allow us to approximate the spatially non-uniform material parameters inside the NP, $\varepsilon_m(T(\bf r))$ and $\kappa_m(T(\bf r))$, by their values on the NP surface, $\varepsilon_m(T_{\textrm{NP}})$ and $\kappa_m(T_{\textrm{NP}})$, where $T_{\textrm{NP}}$ denotes the surface temperature of the NP. Next, we replace the spatially non-uniform absorbed power density in Eq.~(\ref{eq:poissoneq}) by its volume average~\cite{thermo-plasmonics-basics}, $\bar{p}_{\textrm{abs}}(\omega;T_{\textrm{NP}}) \equiv \dfrac{3}{4 \pi a^3} \displaystyle \int \dfrac{\omega\varepsilon_0}{2} \varepsilon_m^{\prime\prime}(\omega,T_{\textrm{NP}}) |{\bf E}(\omega,{\bf r})|^2 d^3 r$; it is more convenient to refer to this expression as $\dfrac{3}{4 \pi a^3} C_{\textrm{abs}}(\omega,T_{\textrm{NP}}) I_{\textrm{inc}}$, where $C_{\textrm{abs}}$ is the absorption cross-section calculated with the uniform metal permittivity $\varepsilon_m(T_{\textrm{NP}})$ and $I_{inc}$ is the intensity of the incoming illumination. Then, one can obtain an approximate analytic solution of Eq.~(\ref{eq:poissoneq}), namely, 
\begin{align}\label{eq:temp_sol}
\begin{cases}
T(r) = T_{\textrm{NP}} + \dfrac{\bar{p}_{\textrm{abs}}(\omega;T_{\textrm{NP}})a^2}{6\kappa_m}\left(1-\dfrac{r^2}{a^2}\right)& \quad\textrm{for}, \quad r < a, \\
\displaystyle\int_{T_{h,0}}^{T(r)}\kappa_h(T) dT = \dfrac{ \bar{p}_{\textrm{abs}}(\omega;T_{\textrm{NP}})a^3 }{3 r}, & \quad\textrm{for}\quad r \geqslant a.
\end{cases}
\end{align}
In Eq.~(\ref{eq:temp_sol}), $T_{\textrm{NP}}$ is now an {\em unknown} variable which needs to be determined by fixed point iterations at $r = a$, namely,
\begin{align}\label{eq:tempnpsc}
\int_{T_{h,0}}^{T_{\textrm{NP}}}\kappa_h(T)dT = \dfrac{ \bar{p}_{\textrm{abs}}(\omega;T_{\textrm{NP}})a^2 }{3} = \dfrac{ C_{\textrm{abs}}(\omega,T_{\textrm{NP}}) I_{\textrm{inc}} }{4 \pi a}.
\end{align}
In contrast to previous studies~\cite{thermo-plasmonics-basics,Hashimoto-nanoscale-cooling,thermo-plasmonics-review}, Eq.~(\ref{eq:tempnpsc}) correctly takes account of the temperature dependence (and hence the spatial dependence) of the host thermal conductivity. In particular, the integration in the left-hand side of Eq.~(\ref{eq:tempnpsc}) indicates that the relative importance of the first-order thermoderivatives of $\kappa_h$ is reduced by a factor of 2, as we shall see later (e.g., in Eq.~(\ref{eq:tempnp_approx}) below). This is because only a small region of the host medium is heated up by the NP.

For weak illumination, one can neglect the temperature dependence of all parameters. In this case, Eq.~(\ref{eq:tempnpsc}) provides a closed form solution for the nanoparticle temperature~\cite{thermo-plasmonics-basics}, 
\begin{align}\label{eq:T_np_lin_sol}
T_{\textrm{NP,I}} = T_{h,0} + \dfrac{C_{\textrm{abs}}(\omega,T_{h,0})I_{\textrm{inc}}}{4\pi\kappa_h a}.
\end{align}
Eq.~(\ref{eq:T_np_lin_sol}) is the linear (first-order) approximation of Eq.~(\ref{eq:tempnpsc}), hence, its solution is denoted by the subscript I.

\section{How to characterize the thermo-optic nonlinearity?}\label{sec:characterize_nlty}
With a formulation for determining the temperature dependence on the incoming intensity in hand, we can now combine this knowledge with the temperature-dependent parameter models (Eq.~(\ref{eq:epsm_2nd_poly}) for the permittivity and the corresponding model for the thermal conductivity of the host) in order to characterize the overall nonlinear thermo-optic response of the metal to the optical illumination. 

Before doing so, one has to clarify two somewhat overlooked complications. First, for materials with a temporally- and spatially-local response (e.g., materials with a pure electronic response), the nonlinear response is conventionally defined via $\epsilon(|\bf{E}|^2)$ where $\bf{E}$ is the local-field and all the relevant coefficients in $\epsilon(|\bf{E}|^2)$ include only intrinsic parameters associated with the NP material. However, in systems in which there is significant contrast between the dielectric constants of the scatterer (the NP, in our case) and the host, the electric field inside the scatterer is strongly affected also by the illumination pattern, the NP geometry and the optical properties of the host. As a result, the nonlinear optical response to a given illumination level is no longer a purely intrinsic property of the metal and can be very sensitive to these properties. This effect contributes to the wide variations of reported nonlinearity values, as seen e.g., through the local-field correction usually introduced to the nonlinear response of small NPs, see, e.g.,~\cite{Hache-cubic-metal-nlty}.

A second complication in the definition of the nonlinear optical response of metal NPs originates from the fact that they have a strong nonlocal thermal response in time~\cite{Stoll_review,Langbein_measure_plasmon_dyn_AuNP_PRB_2012,Biancalana_NJP_2012} and space~\cite{ICFO_Sivan_metal_diffusion,Sivan_Spector_metal_diffusion} which prevents one from being able to directly link the permittivity and local-field at all. Instead, the nonlinearity has to be determined indirectly (e.g., through the dependence of the temperature on the local-field), and now involves {\em additional non-intrinsic properties} such as the thermal properties of the host material~\footnote{This is true for other nonlocal mechanisms, like charge transfer, acoustic response etc.. }. Furthermore, while the temporal nonlocality is of little consequence when the illumination involves just a single temporal frequency (i.e., for the CW illumination case studied in this manuscript), the non-uniformity of the permittivity in space causes the distributions of the electric field and temperature to be quite different - indeed, while the field is, in general, non-uniform across the NP, the temperature is nearly uniform~\cite{thermo-plasmonics-basics,Un-Sivan-size-thermal-effect}. Thus, with the exception of small spherical NPs (see relevant discussion in~\cite{Gurwich-Sivan-CW-nlty-metal_NP}) and potentially also ultrafast illumination, the formulation of the nonlinear optical response as $\epsilon_m(|\bf{E}|^2)$ is simply {\em incorrect}! In that sense, one should refrain from assigning a $\chi^{(3)}$ or $n_2$ values as these are considered to be intrinsic properties of the metal. This frequently ignored fact is the second main source for the wide variability of reported values for the nonlinear optical response of metals.

In light of the above, in what follows we choose instead to characterize the nonlinearity through the dependence of the NP temperature, of the metal permittivity $\varepsilon_m$ and of the scattered power $P_{\textrm{sca}}$ on the {\em incoming intensity} $I_{inc}$. Indeed, one can obtain a direct relation to the incoming intensity (rather than to the local-field) if it consists only of a single spatial frequency component, i.e., a plane wave. More generally, it will be easy to appreciate that since the nonlinearity of metals is so strong, it is ``visible by eye'' when plotted, and can be characterized in percentage with respect to the ambient response, with no need for writing rigorous coefficients for the nonlinearity.

\section{Results}\label{sec:results}
Following the experiments of \cite{plasmonic-SAX-ACS_phot,plasmonic-SAX-PRL,plasmonic-SAX-OE,plasmonic-SAX-rods-Ag,plasmonic-SAX-rods-Ag,Jagadale_SWChu_NL_abs_sca_2019}, we study the thermo-optic nonlinearities of Au and Ag NPs of different sizes immersed in oil (with permittivity $\varepsilon_h = 2.235$, thermal conductivity $\kappa_{h,0} = 0.2873$ W m$^{-1}$ K$^{-1}$ and its thermoderivative $B_{\kappa,h} = 1.297\times 10^{-4}$ W m$^{-1}$ K$^{-2}$). In particular, we will focus on the on-resonance case.

\subsection{Numeric results - Au NPs}\label{sec:num_au}
\begin{figure}[h]
\centering
\includegraphics[width=1\textwidth]{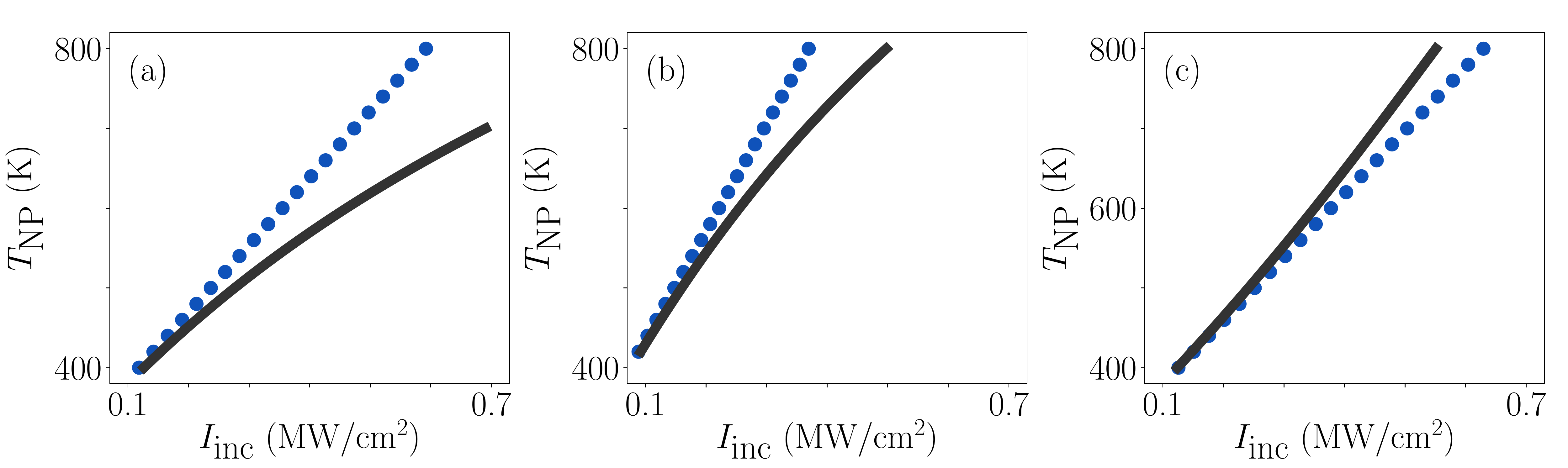}
\caption{\label{fig:T_vs_Iinc_Au} The NP temperature (solution of Eq.~(\ref{eq:tempnpsc}); black solid lines) as a function of the illumination intensity for Au NPs of different sizes and wavelengths. (a) $a = $ 20 nm and $\lambda =$ 550 nm, (b) $a = $ 30 nm and $\lambda =$ 564 nm, and (c) $a = $ 50 nm and $\lambda =$ 610 nm. The linear approximate solutions $\Delta T_{\textrm{NP,I}}$ are shown by blue dotted lines.}
\end{figure}

FIG.~\ref{fig:T_vs_Iinc_Au} shows the solutions of Eq.~(\ref{eq:tempnpsc}) for particle sizes $a = $ 20 nm, 30 nm and 50 nm. For a fair comparison, the wavelengths of the illumination are set to their respective electric dipole resonance wavelengths (550 nm, 564 nm and 610 nm). One can see that for low intensities, the particle temperature increases monotonically with the incoming intensity, with a slope that depends on the particle size, wavelength and material properties (see discussion in~\cite{Un-Sivan-size-thermal-effect}). At higher intensities, this slope changes, depending on the particle size. Specifically, for $a = $ 20 nm and 30 nm, the rate of the temperature increase is sub-linear (i.e., it {\em slows down} as the incoming intensity increases with respect to the low intensity response), in good agreement with the quasi-static calculations in Ref.~\cite{Gurwich-Sivan-CW-nlty-metal_NP}; for $a = 50$ nm, by contrast, the temperature increase rate is super-linear (i.e., it {\em grows} with the incoming intensity due to the increase of absorption with temperature). The latter effect is not captured by the quasi-static approximation used in Ref.~\cite{Gurwich-Sivan-CW-nlty-metal_NP} (see more details in Section~\ref{sec:fur_analy}). For a temperature rise of 400 K, the deviations of the solution~(\ref{eq:tempnpsc}) from the linear (first-order) approximation~(\ref{eq:T_np_lin_sol}) are $\sim -30$\% for $a = 20$ nm, $\sim -15$\% for $a = 30$ nm, and $\sim + 10$\% for $a = 50$ nm. Such a nonlinear response is unusually large for the deep subwavelength scales and moderately high incident intensities involved.

\begin{figure}[h]
\centering
\includegraphics[width=1\textwidth]{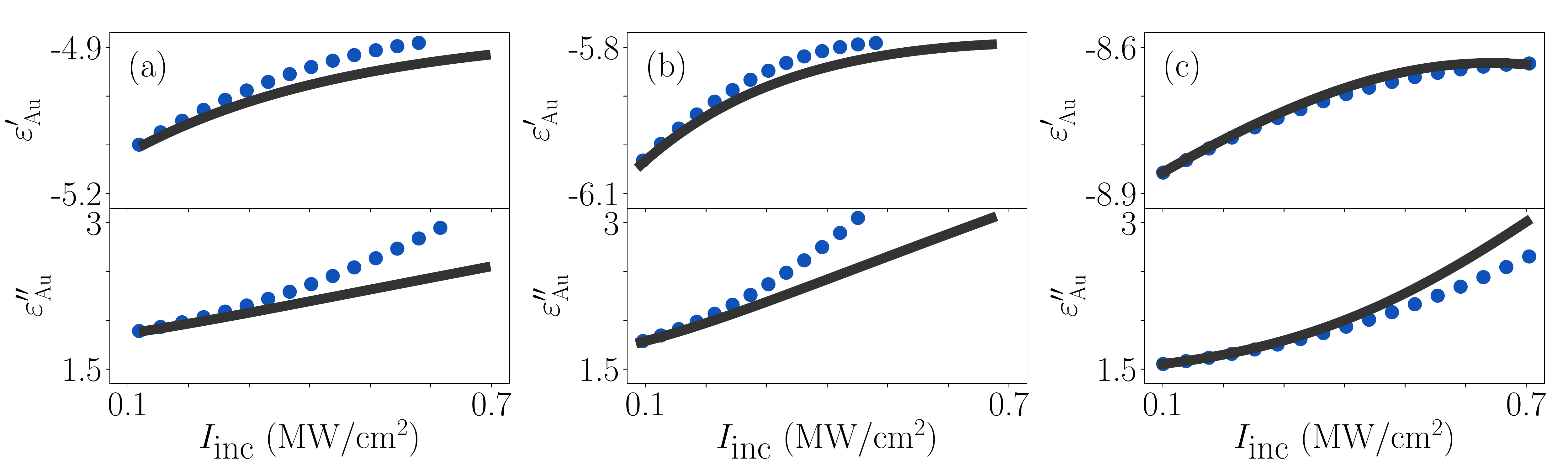}
\caption{\label{fig:epsm_vs_Iinc_Au} Same as Fig.~\ref{fig:T_vs_Iinc_Au} for the permittivity (real part and imaginary part) of Au. }
\end{figure}

Fig.~\ref{fig:epsm_vs_Iinc_Au} shows the corresponding changes of the real and imaginary parts of the Au permittivity. One can see that the relative change of $\varepsilon_{\textrm{Au}}^{\prime}$ is $\sim 5\%$ but the relative change of $\varepsilon_{\textrm{Au}}^{\prime\prime}$ is $25 - 70\%$ depending on the NP size. As mentioned, this is an unusual large nonlinearity for the associated subwavelength scales involved. As noted already in~\cite{Gurwich-Sivan-CW-nlty-metal_NP}, the greater sensitivity of the scattered intensity on the imaginary part of the permittivity is in accord with the experimental findings reported in~\cite{SUSI}.

\begin{figure}[h]
\centering
\includegraphics[width=1\textwidth]{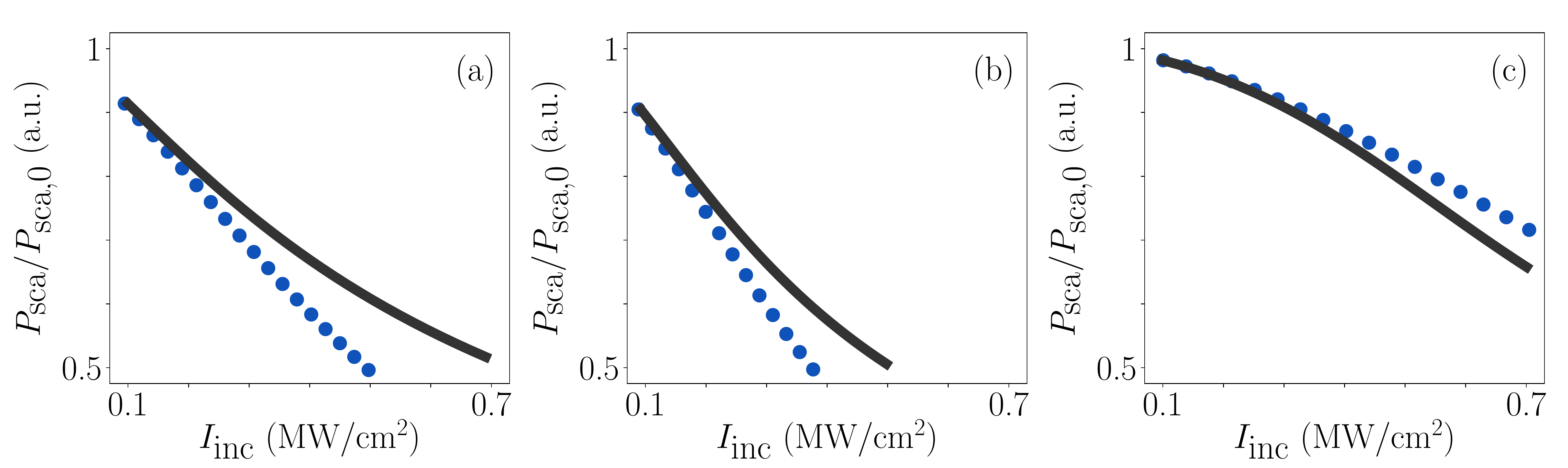}
\caption{\label{fig:Nsca_vs_Iinc_Au} Same as Fig.~\ref{fig:T_vs_Iinc_Au} for the normalized scattered power.}
\end{figure}

In Fig.~\ref{fig:Nsca_vs_Iinc_Au}, we show the corresponding normalized scattered power ($P_{\textrm{sca}}/P_{\textrm{sca,0}}$) from the NP, where $P_{\textrm{sca,0}}$ is the scattered power obtained when the NP is kept at the ambient temperature. One can see that for $a=$ 20 nm and 30 nm, the exact numerical results decrease slower than their first-order approximation $P_{\textrm{sca},I}$ due to the slowing down of the rate of the temperature rise (see Fig.~\ref{fig:T_vs_Iinc_Au}(a) and (b)); for $a=$ 50 nm, the opposite happens because $T_{\textrm{NP}}$ increases faster than $T_{\textrm{NP,I}}$, see Fig.~\ref{fig:T_vs_Iinc_Au}(c). 

\subsection{Numeric results - Ag NPs}\label{sec:num_ag}
\begin{figure}[h]
\centering
\includegraphics[width=1\textwidth]{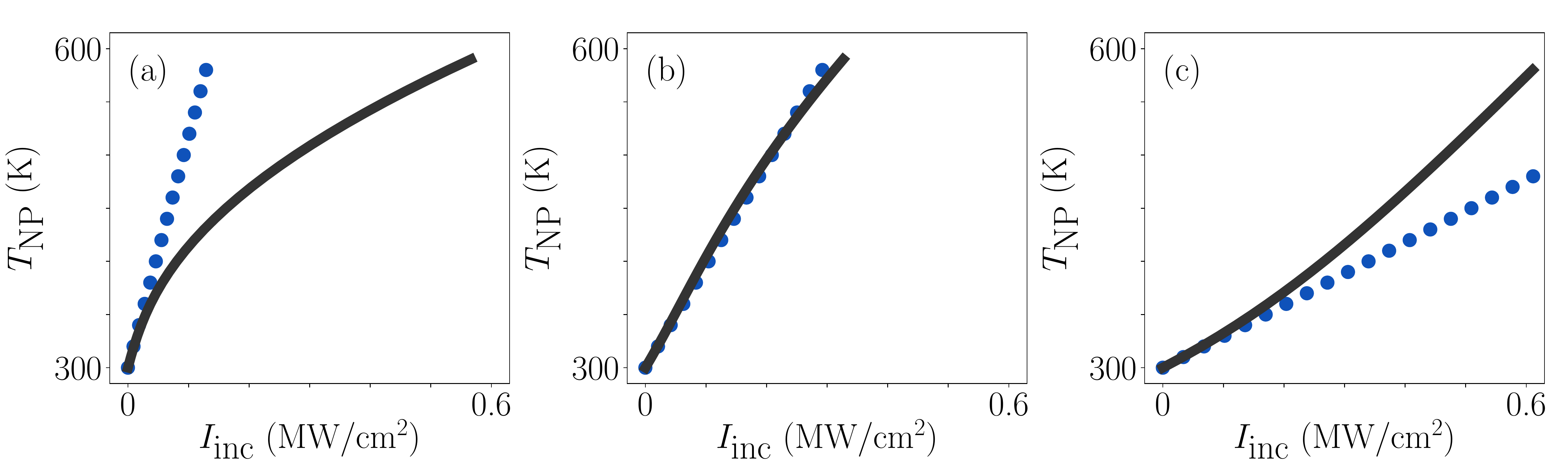}
\caption{\label{fig:T_vs_Iinc_Ag} Same as FIG.~\ref{fig:T_vs_Iinc_Au} for Ag NPs (a) $a = $ 10 nm and $\lambda =$ 420 nm (b) $a = $ 20 nm and $\lambda =$ 435 nm, and (c) $a = $ 30 nm and $\lambda =$ 460 nm. }
\end{figure}

We now replace the Au NP with a Ag NP, while all other conditions remain the same as in Section~\ref{sec:num_au}. FIG.~\ref{fig:T_vs_Iinc_Ag} shows the results of the temperature rise for particle sizes $a =$ 10 nm, 20 nm and 30 nm. Again, we set the wavelengths of the illumination to be their electric dipole resonance wavelengths (420 nm, 435 nm and 460 nm). The results of Ag NPs shown in FIG.~\ref{fig:T_vs_Iinc_Ag} are qualitatively similar to the case of Au NPs (FIG.~\ref{fig:T_vs_Iinc_Au}). The most striking difference between the results shown in FIG.~\ref{fig:T_vs_Iinc_Ag} with respect to Au NPs is the much stronger nonlinearity. 

\begin{figure}[h]
\centering
\includegraphics[width=1\textwidth]{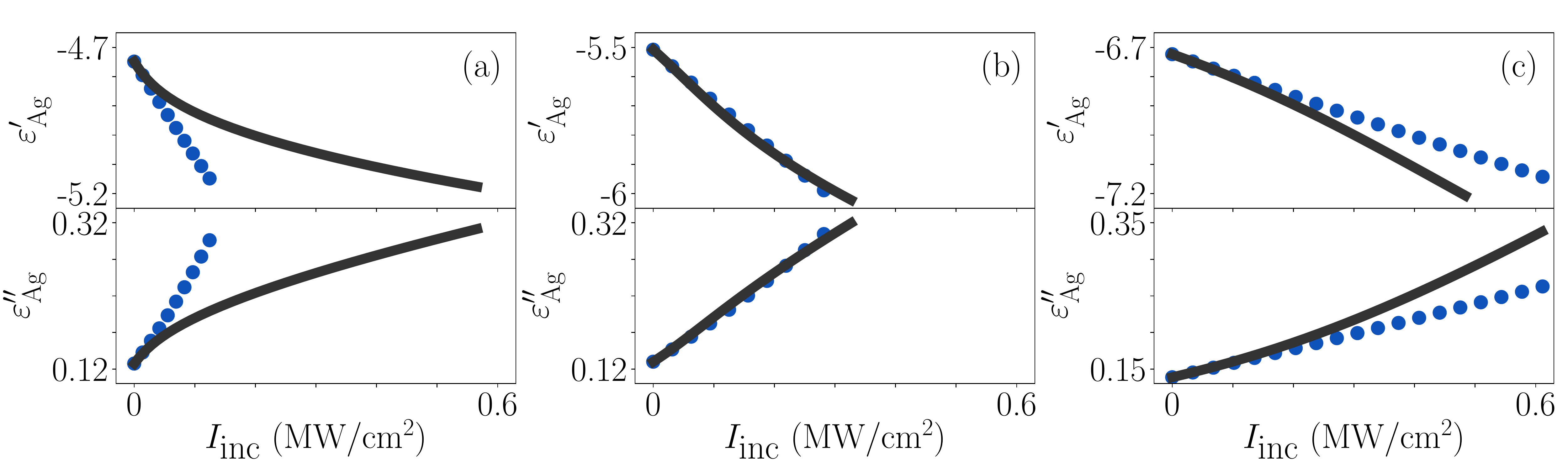}
\caption{\label{fig:epsm_vs_Iinc_Ag} Same as Fig.~\ref{fig:T_vs_Iinc_Ag} for the permittivity (real part and imaginary part) of Ag.}
\end{figure}

For the case of $a = 10$ nm (FIG.~\ref{fig:T_vs_Iinc_Ag}(a)), when the incoming intensity increases from 0 to 0.1 MW/cm$^2$ ($\Delta T_\textrm{NP} < 100$ K), the exact solution deviates from its linear approximation by $\sim -50$\%; when the the incoming intensity is larger than 0.1 MW/cm$^2$ (200 K $< \Delta T_\textrm{NP} < $ 300 K), the slowdown of the temperature becomes even more significant. For the case of $a = 20$ nm (FIG.~\ref{fig:T_vs_Iinc_Ag}(b)), the exact solution coincides with its linear approximation. For the case of $a = 30$ nm (FIG.~\ref{fig:T_vs_Iinc_Ag}(c)), the temperature growth of the exact solution speeds up and deviates from its linear approximation by $\sim 40$\% for a temperature rise of 300 K. Accordingly, the relative change of the $\varepsilon_\textrm{Ag}^{\prime}$ is $\sim 7 - 10$\% and the relative change of $\varepsilon_\textrm{Ag}^{\prime\prime}$ is $\sim 120 - 160$\% depending on the NP size, see FIG.~\ref{fig:epsm_vs_Iinc_Ag}. This, again, indicates a stronger nonlinearity of Ag NPs than that of Au NPs; this is a direct result of the much smaller imaginary part of the Ag permittivity compared to that of Au (see Section~\ref{sec:fur_analy}).

\begin{figure}[h]
\centering
\includegraphics[width=1\textwidth]{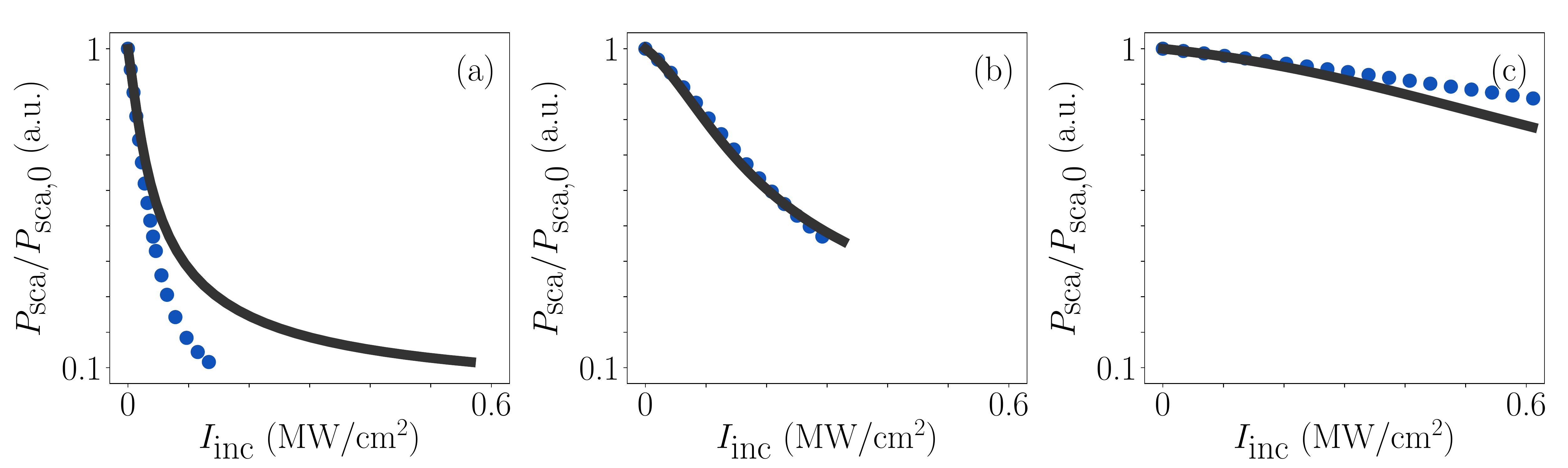}
\caption{\label{fig:Nsca_vs_Iinc_Ag} The normalized scattered power from Ag NPs as a function of the illumination intensity. The particle sizes and the wavelengths are the same as FIG.~\ref{fig:T_vs_Iinc_Ag}.}
\end{figure}

The behaviour of the scattered power from the Ag NPs is qualitatively the same as for Au (See FIG.~\ref{fig:Nsca_vs_Iinc_Au} and FIG.~\ref{fig:Nsca_vs_Iinc_Ag}) except that the nonlinear behaviour is quantitatively stronger, as for the temperature rise and the permittivity change. 

\section{Analysis}\label{sec:fur_analy}
\subsection{Analysis of the temperature and of the nonlinear response and $\epsilon_m(\Delta T)$}
The nonlinear response of the single spherical NP can be characterized via the permittivity $\varepsilon_m(T_{NP}(I_{\textrm{inc}}))$. Thus, it has an intensity-dependence similar to that of $\Delta T_{NP}$, so that we can characterize both simultaneously. Notably, due to the implicit nature of Eq.~(\ref{eq:tempnpsc}), it is difficult to extract physical insights from its self-consistent solution. Therefore, in order to obtain such insights, we approximate Eq.~(\ref{eq:tempnpsc}) by a polynomial expansion in $\Delta T_{\textrm{NP}}\equiv T_{\textrm{NP}} - T_{h,0}$. It turns out that this approximation yields quite satisfactory results even for the maximal heating range of $\Delta T_{\textrm{NP}} \approx 400$ K (not shown). This approximation allows us to employ the Pad\'e approximants~\cite{Sihvola_pade_to_mie_PhysRevB.94.140301,Sihvola_pade_to_mie_review_2017} so that the absorption cross-section is written as a rational function of $x \equiv ka = 2\pi\sqrt{\varepsilon_h}a/\lambda$ (see details in Ref.~\cite[ Appendix B]{Un-Sivan-size-thermal-effect}), namely,
\begin{align}\label{eq:cabs_pd}
C_{\textrm{abs}}(\omega,T_{\textrm{NP}}) \cong \dfrac{12\pi}{k^2} \dfrac{\varepsilon_m^{\prime\prime}(\omega,T_{\textrm{NP}}) \varepsilon_h(\omega) x^3(1 + 2x^2/5)}{Q(\varepsilon_m(\omega,T_{\textrm{NP}}),\varepsilon_h(\omega),x)},
\end{align}
where 
\begin{align}\label{eq:Q}
Q(\varepsilon_m(\omega,T),\varepsilon_h(\omega),x) = q_1^2(\varepsilon_m(\omega,T),\varepsilon_h(\omega),x) + q_2^2(\varepsilon_m(\omega,T),\varepsilon_h(\omega),x),
\end{align}
and where $q_1$ and $q_2$ are, respectively, given by $q_1(\varepsilon_m(\omega,T),\varepsilon_h(\omega),x) = \varepsilon_m^{\prime}(\omega,T)\left(1-\dfrac{3}{5}x^2\right) + 2\varepsilon_h(\omega)\left(1 + \dfrac{3}{5}x^2\right) + \dfrac{2}{3}\varepsilon_m^{\prime\prime}(\omega,T)x^3$ and $q_2(\varepsilon_m(\omega,T),\varepsilon_h(\omega),x) = \varepsilon_m^{\prime\prime}(\omega,T) \left(1-\dfrac{3}{5}x^2\right) - \dfrac{2}{3}( \varepsilon_m^{\prime}(\omega,T) - \varepsilon_h(\omega))x^3$. The electric dipole resonance condition is determined by $q_1 = 0$ and the quality factor of the resonance is quantified by $q_2$. The $x^2$ and $x^3$ terms are, respectively, recognized as the dynamic depolarization~\cite{meier_dyn_depolar_1983} and radiation damping~\cite{Wokaun_rad_damp_PhysRevLett.48.957} effects. 

A few points can be noted already. First, the change of the value of $\varepsilon_m^{\prime\prime}$ has two competing effects on the absorption cross-section: on one hand, the $\varepsilon_m^{\prime\prime}$ in the numerator of Eq.~(\ref{eq:cabs_pd}) (hence, the absorption itself) increases linearly with $\varepsilon_m^{\prime\prime}$; on the other hand, the $\varepsilon_m^{\prime\prime}$ in the denominator of Eq.~(\ref{eq:cabs_pd}) causes the value of $Q$ to increase with $\varepsilon_m^{\prime\prime}$, thus, it causes the resonance quality factor (hence, the absorption) to decrease as well (see the definition of $Q$ in Eq.~(\ref{eq:Q})). However, by how much the resonance quality factor (hence, the absorption) is reduced depends on the NP size. Specifically, when the NP size is small, $q_2 \rightarrow \varepsilon_m^{\prime\prime}$, so that the absorption cross-section is inversely proportional to $\varepsilon_m^{\prime\prime}$ and decreases with the temperature rise; on the other hand, when the NP size is large enough (such that the $x^3$ term dominates), $q_2$ (hence, the quality factor) becomes much more weakly-dependent on $\varepsilon_m^{\prime\prime}$, so that the absorption cross-section becomes proportional to $\varepsilon_m^{\prime\prime}$ (through the numerator) and increases with the temperature rise; we shall see this explicitly later. 

Second, at resonance, the change of $\varepsilon_m^{\prime}$ generically causes a decrease of the absorption due to a shift of resonance wavelength (such that $|q_1|$ increases). It has been shown in~\cite{Gurwich-Sivan-CW-nlty-metal_NP} that when the temperature rise is small, the effect of this resonance shift is secondary for sufficiently small NPs. As we show below, the resonance shift is even smaller for large NPs, yet, when the temperature rise is higher, the effect of the resonance shift can be significant, especially for small metal NPs with low-loss, e.g. Ag.

With the assumptions described in Section~\ref{sec:config} and the Pad\'e approximations, the NP temperature~(\ref{eq:tempnpsc}) can be rewritten as an implicit equation in $\Delta T_{\textrm{NP}}$, namely,
\begin{equation}\label{eq:tempnp_approx}
\Delta T_{\textrm{NP}}\left(1 + \dfrac{B_{\kappa,h}}{2\kappa_{h,0}}\Delta T_{\textrm{NP}}\right)\left\{\dfrac{1 + \dfrac{Q_1}{Q_0}\Delta T_{\textrm{NP}} + \dfrac{Q_2}{Q_0}\Delta T_{\textrm{NP}}^2 + \cdots}{1 + \dfrac{B_m^{\prime\prime}}{\varepsilon_{m,0}^{\prime\prime}}\Delta T_{\textrm{NP}} + \dfrac{D_m^{\prime\prime}}{\varepsilon_{m,0}^{\prime\prime}}\Delta T_{\textrm{NP}}^2}\right\} = \Delta T_\textrm{NP,I}, 
\end{equation}
where $Q_0 = Q(\varepsilon_{m,0}(\omega),\varepsilon_{h}(\omega),x)$, $Q_1 = \dfrac{\partial Q}{\partial T}\bigg|_{T_{h,0}}$ and $Q_2 = \dfrac{1}{2} \dfrac{\partial^2 Q}{\partial T^2}\bigg|_{T_{h,0}}$ are the Taylor expansion coefficients of $Q$. Eq.~(\ref{eq:tempnp_approx}) is a generalization of Eq.~(4) of Ref.~\cite{Gurwich-Sivan-CW-nlty-metal_NP} for NPs more than a few nm in size (note the similarity of notations). Specifically, under the quasi-static approximation ($x \rightarrow 0$), $q_1 \rightarrow \varepsilon_m^{\prime} + 2\varepsilon_h$ and $q_2 \rightarrow \varepsilon_m^{\prime\prime}$. In this case, $Q_0 \rightarrow (\varepsilon_{m,0}^{\prime} + 2\varepsilon_h)^2+\varepsilon_{m,0}^{\prime\prime 2}$, $Q_1 \rightarrow 2(\varepsilon_{m,0}^{\prime} + 2\varepsilon_h)B_m^{\prime} + 2\varepsilon_{m,0}^{\prime\prime}B_m^{\prime\prime}$ and $Q_2 \rightarrow B_m^{\prime 2} + B_m^{\prime\prime 2}$. After some lengthy algebra, one can verify that Eq.~(\ref{eq:tempnp_approx}) reduces to the fourth-order polynomial equation in $\Delta T_{\textrm{NP}}$ appearing in Eq.~(4) of Ref.~\cite{Gurwich-Sivan-CW-nlty-metal_NP}. 

We notice that Eq.~(\ref{eq:tempnp_approx}) has the form $\Delta T_{\textrm{NP,I}}(\Delta T_{\textrm{NP}})$ rather than the more desired form $\Delta T_{\textrm{NP}}(\Delta T_{\textrm{NP,I}}(I_{\textrm{inc}}))$. This can be fixed by applying the Lagrange inversion theorem to Eq.~(\ref{eq:tempnp_approx}) so that $\Delta T_{\textrm{NP}}$ can be expanded into a sum of a power series in $I_{\textrm{inc}}$. However, we find that (not shown) the resulting expansion of $\Delta T_{\textrm{NP}}(I_{\textrm{inc}})$ converges to the exact solution Eq.~(\ref{eq:tempnpsc}) only for a small range of values of $\Delta T_{\textrm{NP}}$, especially for small metal NPs with low losses; for example, for an Ag NP of 10 nm in radius, it converges only when $\Delta T_{\textrm{NP}} < 100$ K. Moreover, in the more desirable expansion $\Delta T_{\textrm{NP}}(I_{\textrm{inc}})$, the physical effects are mixed in the coefficients, making it difficult to extract meaningful physical insights. In contrast, the solution of Eq.~(\ref{eq:tempnp_approx}) turns out to be nearly indistinguishable from the exact solution~(\ref{eq:tempnpsc}) even for large temperature rises. In addition, relation~(\ref{eq:tempnp_approx}) provides direct physical insights into the origins of the nonlinear response. Specifically, the first factor $\Delta T_{\textrm{NP}}$ in Eq.~(\ref{eq:tempnp_approx}) represents the linear temperature response when the intensity (and hence $\Delta T_{\textrm{NP}}$) is small; the second factor $\left(1 + \dfrac{B_{\kappa,h}}{2\kappa_{h,0}}\Delta T_{\textrm{NP}}\right)$ corresponds to the nonlinearity associated with the temperature-dependent host thermal conductivity (i.e., it can be traced back to the term $\int_{T_{h,0}}^{T_{\textrm{NP}}} \kappa_h(T)/\kappa_h(T_{h,0})dT$ in Eq.~(\ref{eq:tempnpsc})). This effect of the host thermal conductivity is disentangled from the other physical effects. The terms in the curly brackets in Eq.~(\ref{eq:tempnp_approx}) refer to the contributions from the temperature-dependent metal permittivity of the NP. For these reasons, we study the less conventional but more accurate and physically meaningful Eq.~(\ref{eq:tempnp_approx}) to provide a more detailed analysis of the numerical results above. 

In order to further elucidate the relative importance of the various physical effects on the NP temperature, we expand the quotient in the curly brackets in Eq.~(\ref{eq:tempnp_approx}) in a Taylor series of $\Delta T_{\textrm{NP}}$, so that Eq.~(\ref{eq:tempnp_approx}) can be rewritten as a sum of a power series in $\Delta T_{\textrm{NP}}$. It can be shown that all the coefficients of this power series can be expressed in terms of the normalized thermoderivatives of the absorption cross-section~\footnote{One can verify this identification by directly differentiating Eq.~(\ref{eq:cabs_pd}) with respective to $T$ using the chain rule. It can also be shown that this expansion is equivalent to the Taylor series of Eq.~(\ref{eq:tempnpsc}) with respect to $\Delta T_{\textrm{NP}}$.}, e.g. $\left. \dfrac{1}{C_{\textrm{abs}}}\dfrac{dC_{\textrm{abs}}}{dT}\right|_{T_{h,0}}$, $\left. \dfrac{1}{C_{\textrm{abs}}}\dfrac{d^2C_{\textrm{abs}}}{dT^2}\right|_{T_{h,0}}, \cdots$ (see details in Appendix~\ref{app:detail_analys}), namely, 
\begin{align}\label{eq:tempnp_fur_approx}
& \Delta T_{\textrm{NP}}\left(1 + \dfrac{B_{\kappa,h}}{2\kappa_{h,0}}\Delta T_{\textrm{NP}}\right) \left[1 + \left(\dfrac{Q_1}{Q_0} - \dfrac{B_m^{\prime\prime}}{\varepsilon_{m,0}^{\prime\prime}}\right)\Delta T_{\textrm{NP}} \right.\nonumber \\
&\left. + \left(\dfrac{Q_2}{Q_0} + \left(\dfrac{B_m^{\prime\prime}}{\varepsilon_{m,0}^{\prime\prime}}\right)^2-\dfrac{Q_1}{Q_0}\dfrac{B_m^{\prime\prime}}{\varepsilon_{m,0}^{\prime\prime}} - \dfrac{D_m^{\prime\prime}}{\varepsilon_{m,0}^{\prime\prime}}\right)\Delta T_{\textrm{NP}}^2 + \mathcal{O}(\Delta T_{\textrm{NP}}^3)\right] \nonumber \\
&= \Delta T_{\textrm{NP}} \left(1 + \dfrac{B_{\kappa,h}}{2\kappa_{h,0}}\Delta T_{\textrm{NP}}\right)\left\{ 1 - \left(\dfrac{1}{C_{\textrm{abs}}}\dfrac{dC_{\textrm{abs}}}{dT}\right)_{T_{h,0}}\Delta T_{\textrm{NP}}\right. \nonumber \\
&- \left.\dfrac{1}{2}\left[\dfrac{1}{C_{\textrm{abs}}}\dfrac{d^2C_{\textrm{abs}}}{dT^2} - 2\left(\dfrac{1}{C_{\textrm{abs}}}\dfrac{dC_{\textrm{abs}}}{dT}\right)^2\right]_{T_{h,0}}\Delta T_{\textrm{NP}}^2 + \mathcal{O}(\Delta T_{\textrm{NP}}^3)\right\} = \Delta T_{\textrm{NP,I}}.
\end{align} 
As one could expect, the solution of Eq.~(\ref{eq:tempnp_fur_approx}) converges to the exact solution Eq.~(\ref{eq:tempnpsc}) for a moderately high temperature rise only (not shown). Therefore, in the following, we choose to analyze the coefficients instead of the solution of Eq.~(\ref{eq:tempnp_fur_approx}) itself. We also examine how each of the thermoderivatives affects the NP temperature and study the dependence of the nonlinear response on the NP size.

\subsubsection{The effects of the thermoderivatives of the permittivity on the temperature rise}
In this subsection, we focus on the coefficients in the curly brackets in Eq.~(\ref{eq:tempnp_fur_approx}) and examine the effects of the thermoderivatives of the permittivity on the absorption cross-section and on the temperature rise. Applying the chain rule to the coefficient  $\left(\dfrac{1}{C_{\textrm{abs}}}\dfrac{dC_{\textrm{abs}}}{dT}\right)_{T_{h,0}}$, one can see that the relative importance of $B_m^{\prime}$ and $B_m^{\prime\prime}$ is weighted by the sensitivity of $C_{\textrm{abs}}$ to $\varepsilon_m^{\prime}$ and to $\varepsilon_m^{\prime\prime}$, i.e., 
\begin{align}\label{eq:cabs_sen}
\left(\dfrac{1}{C_{\textrm{abs}}}\dfrac{dC_{\textrm{abs}}}{dT}\right)_{T_{h,0}} = \left(\dfrac{1}{C_{\textrm{abs}}}\dfrac{\partial C_{\textrm{abs}}}{\partial \varepsilon_m^{\prime}}\right)_{T_{h,0}}B_m^{\prime} + \left(\dfrac{1}{C_{\textrm{abs}}}\dfrac{\partial C_{\textrm{abs}}}{\partial \varepsilon_m^{\prime\prime}}\right)_{T_{h,0}}B_m^{\prime\prime}.
\end{align}
A similar expression has already been used to study the ultrafast nonlinearity of single NPs in~\cite{Del-Fatti-PRL-UF-NLTY-MNR-2011,Stoll_review,Stoll_environment}, even though for a fixed nanoparticle size; these studies showed that $\partial C_{\textrm{ext}}/\partial \varepsilon_m^{\prime}$ and $\partial C_{\textrm{ext}}/\partial \varepsilon_m^{\prime\prime}$ are strongly enhanced at resonance such that they dictate the nonlinear response of the nanoparticle around its plasmonic resonance wavelength. It was also found that the wavelength-dependence of $\partial C_{\textrm{ext}}/\partial \varepsilon_m^{\prime}$ has a Lorentzian-like profile similar to $\partial C_{\textrm{ext}}/\partial \lambda$ and always crosses the horizontal axis near the plasmonic resonance wavelength~\footnote{The extinction cross-section ($C_{\textrm{ext}} = C_{\textrm{sca}} + C_{\textrm{abs}}$) were shown to be dominated by the absorption cross-section for sufficiently small NPs~\cite{Bohren-Huffman-book}.}. 

\begin{figure}[h]
\centering
\includegraphics[width=1\textwidth]{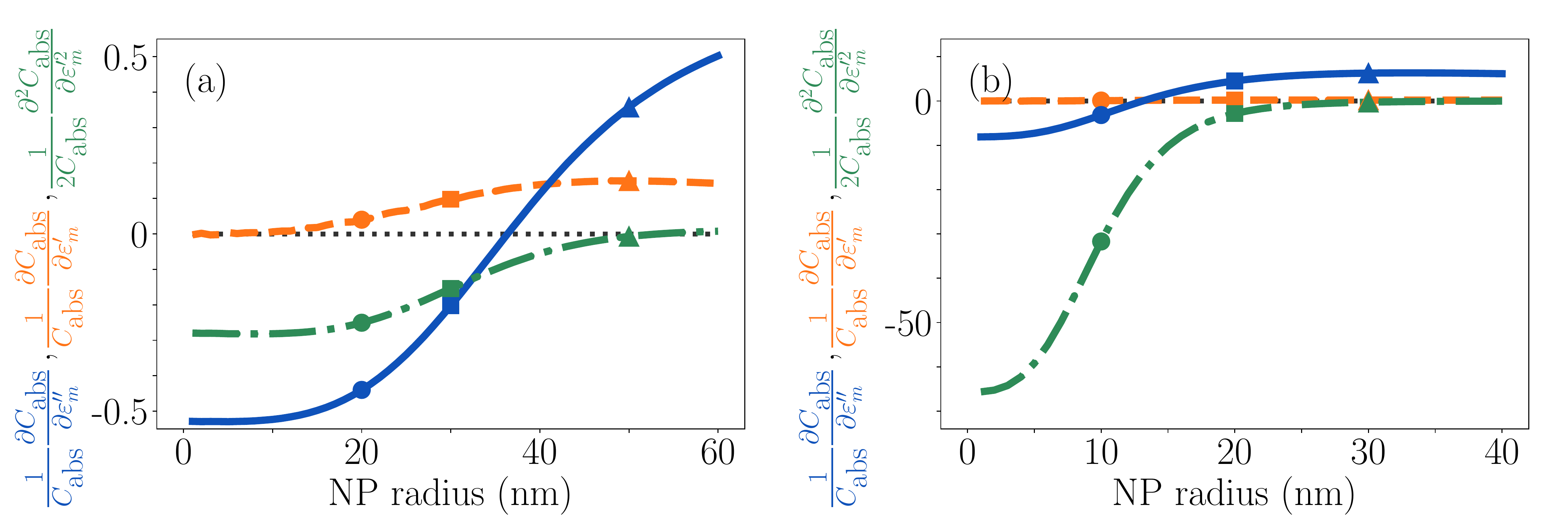}
\caption{\label{fig:cabs_sen_Au_Ag} (Color online) The normalized derivatives of the absorption cross-section respect to $\varepsilon_m^{\prime}$ (orange dashed line: $\dfrac{1}{C_{\textrm{abs}}}\dfrac{\partial C_{\textrm{abs}}}{\partial \varepsilon_m^{\prime}}$ and green dash-dotted line: $\dfrac{1}{2C_{\textrm{abs}}}\dfrac{\partial^2 C_{\textrm{abs}}}{\partial \varepsilon_m^{\prime 2}}$) and to $\varepsilon_m^{\prime\prime}$ (blue solid line: $\dfrac{1}{C_{\textrm{abs}}}\dfrac{\partial C_{\textrm{abs}}}{\partial \varepsilon_m^{\prime\prime}}$) for (a) Au and (b) Ag NPs of different sizes, all evaluated at $T_{h,0}$. The circle, square and triangle dots in (a) and (b) label the cases shown in FIG.~\ref{fig:T_vs_Iinc_Au} and FIG.~\ref{fig:T_vs_Iinc_Ag}, respectively.}
\end{figure}

We now would like to add insights to the results of Refs.~\cite{Del-Fatti-PRL-UF-NLTY-MNR-2011,Stoll_review,Stoll_environment} by exploring the size dependence of the nonlinear response. To do that, we calculate the normalized derivatives of $C_{\textrm{abs}}$ with respect to $\varepsilon_m^{\prime}$ and $\varepsilon_m^{\prime\prime}$ under the electric dipole resonance condition at $T_{h,0}$, i.e., when the resonance wavelength depends on the NP size via $q_1(\varepsilon_{m,0}(\omega),\varepsilon_h(\omega), ka) = 0$. Under these conditions, we get
\begin{subequations}
\begin{align}\label{eq:cabs_sen_pd_real}
\left(\dfrac{1}{C_{\textrm{abs}}}\dfrac{\partial C_{\textrm{abs}}}{\partial \varepsilon_m^{\prime}}\right)_{\substack{q_1 = 0,\\T_\textrm{NP} = T_{h,0}}} = \dfrac{4}{3}\dfrac{x^3}{q_2(\varepsilon_{m,0},\varepsilon_h,x)},
\end{align}
\begin{align}\label{eq:cabs_sen_pd_imag}
\left(\dfrac{1}{C_{\textrm{abs}}}\dfrac{\partial C_{\textrm{abs}}}{\partial \varepsilon_m^{\prime\prime}}\right)_{\substack{q_1 = 0,\\T_\textrm{NP} = T_{h,0}}} = -\dfrac{1}{\varepsilon_{m,0}^{\prime\prime}}\left[1-\dfrac{4}{3}\dfrac{|\varepsilon_{m,0}^{\prime}-\varepsilon_h|}{q_2(\varepsilon_{m,0},\varepsilon_h,x)}x^3\right].
\end{align}
\end{subequations}
The size-dependence of these terms for Au and Ag is shown in FIG~\ref{fig:cabs_sen_Au_Ag}. We can now study these coefficients in two ranges of particle sizes.

{\bf Small NPs (i.e., $x \ll 1$, $a <  20$ nm for Au NPs; $a <  10$ nm for Ag NPs).} In this case, $\left(\dfrac{1}{C_{\textrm{abs}}}\dfrac{\partial C_{\textrm{abs}}}{\partial \varepsilon_m^{\prime\prime}}\right)_{\substack{q_1 = 0,\\ T_\textrm{NP}=T_{h,0}}} \rightarrow - \dfrac{1}{\varepsilon_{m,0}^{\prime\prime}} < 0$; this occurs because of the increase of $\varepsilon_m^{\prime\prime}$ with the temperature causes a reduction of the resonance quality factor. On the other hand, $\left(\dfrac{1}{C_{\textrm{abs}}}\dfrac{\partial C_{\textrm{abs}}}{\partial \varepsilon_m^{\prime}}\right)_{\substack{q_1 = 0,\\T_\textrm{NP}=T_{h,0}}} \rightarrow 0$, so that $B_m^{\prime}$ has a negligible effect on the temperature rise, in agreement with previous report on the weak (second-order) effect of $\varepsilon_m^{\prime}$ on the temperature~\cite{Gurwich-Sivan-CW-nlty-metal_NP}. In this case, Eq.~(\ref{eq:cabs_sen}) reduces to  $\left(\dfrac{1}{C_{\textrm{abs}}}\dfrac{dC_{\textrm{abs}}}{dT}\right)_{\substack{q_1 = 0,\\T_\textrm{NP} = T_{h,0}}} \approx -\dfrac{B_m^{\prime\prime}}{\varepsilon_{m,0}^{\prime\prime}}$. 

However, the effect of $B_m^{\prime}$ on the temperature rise becomes non-negligible beyond the perturbative regime studied in~\cite{Gurwich-Sivan-CW-nlty-metal_NP}, i.e., for temperature rise of more than about 100K. Indeed, when the temperature rise is moderately high ($\Delta T_\textrm{NP} > 200$ K for Au NPs and $\Delta T_\textrm{NP} > 100$ K for Ag NPs), the coefficient of $\Delta T_\textrm{NP}^2$ in the curly bracket in Eq.~(\ref{eq:tempnp_fur_approx}) can be rewritten by \begin{align}\label{eq:dTNPsq_coeff_small_x}
&-\dfrac{1}{2}\left[\dfrac{1}{C_{\textrm{abs}}}\dfrac{d^2 C_{\textrm{abs}}}{dT^2} - 2\left( \dfrac{1}{C_{\textrm{abs}}}\dfrac{d C_{\textrm{abs}}}{dT} \right)^2\right]_{\substack{q_1 = 0, \\ T_\textrm{NP} = T_{h,0}}}\\ \xlongequal{x\rightarrow0} &-\dfrac{1}{2}\left(\dfrac{1}{C_{\textrm{abs}}}\dfrac{\partial^2 C_{\textrm{abs}}}{\partial\varepsilon_m^{\prime\ 2}}\right)_{\substack{q_1 = 0, \\ T_\textrm{NP} = T_{h,0}}}B_m^{\prime\ 2}-\left(\dfrac{1}{C_{\textrm{abs}}}\dfrac{\partial^2 C_{\textrm{abs}}}{\partial\varepsilon_m^{\prime\ 2}}\right)_{\substack{q_1 = 0, \\ T_\textrm{NP}=T_{h,0}}}D_m^{\prime\prime} = \left[ \left(\dfrac{B_m^\prime}{\varepsilon_{m,0}^{\prime\prime}}\right)^2+\dfrac{D_m^{\prime\prime}}{\varepsilon_{m,0}^{\prime\prime}}\right].\nonumber
\end{align}
The first term $(B_m^{\prime}/\varepsilon_{m,0}^{\prime\prime})^2$ in Eq.~(\ref{eq:dTNPsq_coeff_small_x}) causes the absorption cross-section to decrease with the NP temperature via a shift away from resonance, regardless of the sign of $B_m^{\prime}$. This further slows down the temperature rise. Notably, the normalization by $\varepsilon_{m,0}^{\prime\prime}$ (rather than by $\varepsilon_{m,0}^{\prime}$) causes the resonance shift to be stronger for NPs with lower loss.

{\bf Large NPs (i.e., when the $x^3$ term dominates; specifically, for $a > 40$ nm for Au NPs and $a > 30$ nm for Ag NPs).} Different from the case of small NPs, the resonance quality ($\sim q_2$) is weakly-dependent on $\varepsilon_m^{\prime\prime}$ so that $C_\textrm{abs} \sim \varepsilon_m^{\prime\prime}$ (rather than to its inverse, as noted in Eq.~(\ref{eq:cabs_pd})). As a result, $\left(\dfrac{1}{C_{\textrm{abs}}}\dfrac{\partial C_{\textrm{abs}}}{\partial \varepsilon_m^{\prime\prime}}\right)_{\substack{q_1 = 0,\\T_\textrm{NP}=T_{h,0}}} \rightarrow + \dfrac{1}{\varepsilon_{m,0}^{\prime\prime}} > 0$ (instead of $-1/\varepsilon_{m,0}^{\prime\prime}$). This can also verified from Eq.~(\ref{eq:cabs_sen_pd_imag}). Another difference is that $\left(\dfrac{1}{C_{\textrm{abs}}}\dfrac{\partial C_{\textrm{abs}}}{\partial \varepsilon_m^{\prime}}\right)_{\substack{q_1 = 0,\\T_\textrm{NP}=T_{h,0}}} \rightarrow \dfrac{2}{\left|\varepsilon_{m,0}^{\prime} - \varepsilon_h\right|}$, i.e., it does not vanish anymore, so that the effect of $B_m^{\prime}$ on the temperature rise is non-negligible. 

Furthermore, in analogy to Eq.~(\ref{eq:dTNPsq_coeff_small_x}), the coefficient of $\Delta T_\textrm{NP}^2$ in the curly bracket in Eq.~(\ref{eq:tempnp_fur_approx}) becomes (see Appendix~\ref{app:detail_analys}) 
\begin{align}\label{eq:dTNPsq_coeff_large_x}
&-\dfrac{1}{2}\left[\dfrac{1}{C_{\textrm{abs}}}\dfrac{d^2 C_{\textrm{abs}}}{dT^2} - 2\left( \dfrac{1}{C_{\textrm{abs}}}\dfrac{d C_{\textrm{abs}}}{dT} \right)^2\right]_{\substack{q_1 = 0, \\ T_\textrm{NP} = T_{h,0}}}\nonumber\\ \approx &\  \left(\dfrac{B_m^{\prime}}{|\varepsilon_{m,0} - \varepsilon_h|}+\dfrac{B_m^{\prime\prime}}{\varepsilon_{m,0}^{\prime\prime}}\right)^2 - \dfrac{D_m^{\prime}}{|\varepsilon_{m,0}^{\prime} - \varepsilon_h|} - \dfrac{D_m^{\prime\prime}}{\varepsilon_{m,0}^{\prime\prime}}.
\end{align}
Specifically, the effect of $B_m^{\prime\ 2}$ on the temperature rise is normalized by $|\varepsilon_{m,0} - \varepsilon_h|^2$ so that the effect of the shift away from resonance on this coefficient is much weaker than that in the case of small NPs. 

\subsubsection{Comparison of thermoderivatives of the permittivity and thermal conductivity of the host}

As we can see from Eq.~(\ref{eq:tempnp_approx}) (or from Eq.~(\ref{eq:tempnp_fur_approx})), when the thermoderivative of the host thermal conductivity ($B_\kappa$) is positive (negative), the temperature growth rate decreases (increases) as the NP temperature increases. This host dependence is unique to the thermo-optical response to the CW illumination, i.e., it complements the dependence on the thermo-derivatives of the metal permittivity which dominate the ultrafast response. In the numerical examples above,  $\dfrac{B_{\kappa,h}}{2\kappa_{h,0}} > 0$ for oil. Thus, for the case of small NPs, $\dfrac{B_{\kappa,h}}{2\kappa_{h,0}}$ has the same sign as $-\left(\dfrac{1}{C_\textrm{abs}}\dfrac{\partial C_\textrm{abs}}{\partial \varepsilon_m^{\prime\prime}}\right)_{\substack{q_1 = 0, \\ T_\textrm{NP}=T_{h,0}}} B_m^{\prime\prime} \approx \dfrac{B_m^{\prime\prime}}{\varepsilon_{m,0}^{\prime\prime}}$ and the resonance shift effect ($\sim (B_m^{\prime}/\varepsilon_{m,0}^{\prime})^2$), so that these three effects act jointly to cause a substantial slowdown of the temperature growth, see FIG.~\ref{fig:T_vs_Iinc_Au} (a) and FIG.~\ref{fig:T_vs_Iinc_Ag} (a). The slowdown of the temperature growth of small Ag NPs is much more significant than that of small Au NPs because $\varepsilon_\textrm{Ag}^{\prime\prime} \ll \varepsilon_\textrm{Au}^{\prime\prime}$. 

On the other hand, for large NPs, $-\left(\dfrac{1}{C_\textrm{abs}}\dfrac{\partial C_\textrm{abs}}{\partial \varepsilon_m^{\prime\prime}}\right)_{\substack{q_1 = 0, \\ T_\textrm{NP}=T_{h,0}}} B_m^{\prime\prime} \approx -\dfrac{B_m^{\prime\prime}}{\varepsilon_{m,0}^{\prime\prime}}$ is of different sign from $\dfrac{B_{\kappa,h}}{2\kappa_{h,0}}$ and the resonance shift effect ($\sim (B_m^{\prime}/|\varepsilon_{m,0}^{\prime\prime}-\varepsilon_h|)^2$). Thus, the effect of $\dfrac{B_{\kappa,h}}{2\kappa_{h,0}}$ and of ${B_m^{\prime\prime}}/{\varepsilon_{m,0}^{\prime\prime}}$ can counteract each other. This happens for the 40 nm Au NP studied in Section~\ref{sec:num_au}. The slightly faster NP temperature growth rate shown in FIG.~\ref{fig:T_vs_Iinc_Au}(c) is indeed due to the positive value of $B_\textrm{Au}^{\prime}$ used in the simulation and due to the higher-order correction $D_m/\varepsilon_{m,0}^{\prime\prime}$ in Eq.~(\ref{eq:dTNPsq_coeff_large_x}). In contrast, for the example of 30 nm studied in Section~\ref{sec:num_ag}, $B_\textrm{Ag}^{\prime\prime}/\varepsilon_\textrm{Ag,0}^{\prime\prime}$ is much larger than $B_{\kappa,h}/(2\kappa_{h,0})$ and the resonance shift effect so that the temperature growth speeds up substantially. In this case, the speedup of the temperature growth is hardly affected by the temperature dependence of the host thermal conductivity.

\subsection{Analysis of the {\em apparent} nonlinear response}
In order to provide a more complete picture of the nonlinear response, in what follows, we chose to study also another, non-intrinsic yet potentially more accessible observable of the nonlinear response, namely, the nonlinear response of the local-field to the incoming intensity. It is more complicated than the proper nonlinearity (i.e., $\varepsilon_m(T(I_{inc}))$) as it incorporates, again, its own non-trivial dependence on the metal and host permittivities; it is representative also of the intensity dependence of the scattered and absorbed power (not further analyzed).

\begin{figure}[h]
\centering
\includegraphics[width=1\textwidth]{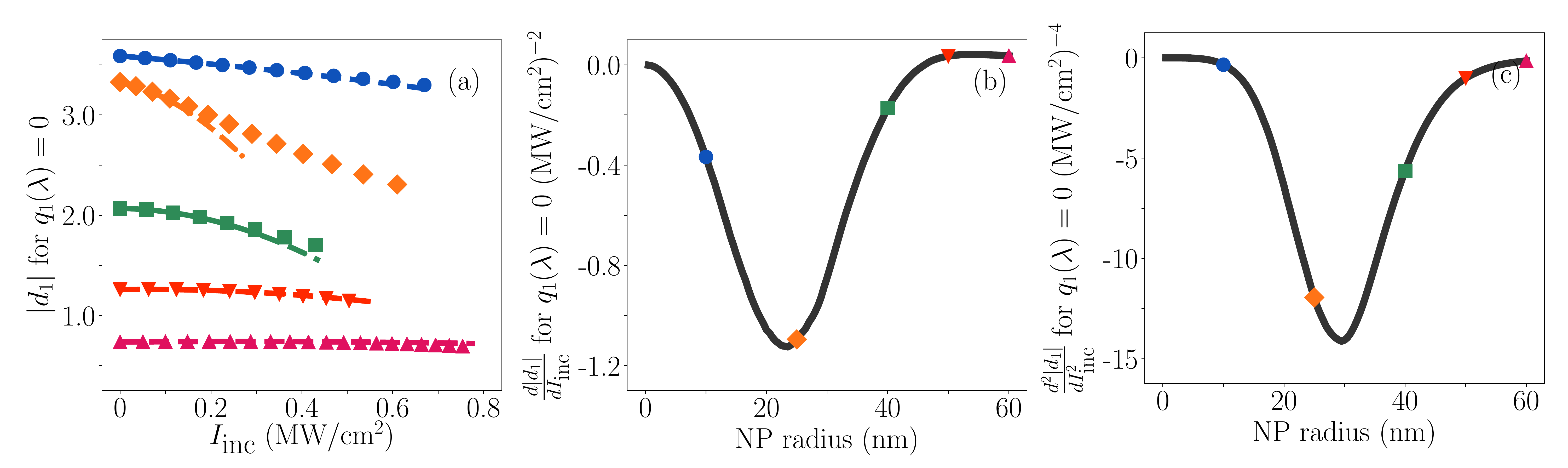}
\caption{(Color online) (a) The Mie coefficient $|d_1|$ for $q_1(\lambda) = 0$ (i.e., when the illumination wavelengths is set to the respective electric dipole resonance wavelength) as a function of the illumination intensity for Au NPs with $a = 10$ nm and $\lambda = 540$ nm (blue circles), $a = 25$ nm and $\lambda = 555$ nm (orange diamonds), $a = 40$ nm and $\lambda = 585$ nm (green squares), $a = 50$ nm and $\lambda = 610$ nm (red down-pointing triangles) and $a = 60$ nm and $\lambda = 640$ nm (magenta up-pointing triangles). The thermoderivatives $B_\textrm{Au}^{\prime}$ and $B_\textrm{Au}^{\prime\prime}$ are positive within this wavelength range, see~\cite{PT_Shen_ellipsometry_gold}. The dashed lines denote the corresponding fits by second-degree polynomials in $I_{\textrm{inc}}$ at $I_{\textrm{inc}} \rightarrow 0$, analogous to an apparent cubic-quintic thermo-optic nonlinearity. (b) The corresponding apparent cubic and (c) the quintic nonlinear coefficients for $q_1(\lambda) = 0$ as a function of $ka$ (black solid line). The markers correspond to the specific cases shown in (a).}\label{fig:d1_Efieldsq}
\end{figure}

For plane wave illumination polarized along $x$ and propagating in the $z$ direction, necessarily, $m = 1$~\cite{Bohren-Huffman-book}. According to Mie theory, the field enhancement is given by ${\bf E}/E_{\textrm{inc}} = \sum\limits_{n=0}^{\infty} i^n\dfrac{2n+1}{n(n+1)}\left(c_n{\bf M}^{(1)}_{o1n} - i d_n{\bf N}^{(1)}_{e1n}\right)$. Here, we use the same notation for the electromagnetic modes of the spherical particle as in Ref.~\cite{Bohren-Huffman-book}. The electric dipole Mie coefficient in the Pad\'e approximation is
\begin{equation}
d_1 = \dfrac{3\left[\varepsilon_h + \frac{1}{10}(\varepsilon_{m,0} + \varepsilon_h)x^2\right] + \frac{3}{10}B_m x^2 \Delta T }{\left\{\splitfrac{q_1(\varepsilon_{m,0},\varepsilon_h,x) + \left[B_m^{\prime}\left(1 - \frac{3}{5}x^2\right) + \frac{2}{3} B_m^{\prime\prime}x^3\right]\Delta T}{+ i q_2(\varepsilon_{m,0},\varepsilon_h,x) +  i\left[B_m^{\prime\prime}\left(1 - \frac{3}{5}x^2\right) - \frac{2}{3} B_m^{\prime}x^3\right]\Delta T}\right\}}.
\end{equation} 
Fig.~\ref{fig:d1_Efieldsq} shows $d_1$ under the condition that $q_1(\varepsilon_{m,0}(\omega),\varepsilon_h(\omega),x) =  0$), i,e., it follows the dipolar resonance (as in Fig.~\ref{fig:cabs_sen_Au_Ag}).

We associate the apparent nonlinear response to the deviation of $d_1$ from its low intensity (i.e., room temperature) value. Thus, the apparent cubic nonlinearity is represented by the first-order derivative of $d_1$ with respect to the incoming intensity. Whenever $d_1$ is nonlinear, the nonlinear thermo-optic response has contributions higher than the cubic description, i.e., the nonlinear response is non-perturbative. The second-order derivative of $d_1$ with respect to the incoming intensity is hence referred to as the quintic nonlinearity. One can see that for small spheres, the local-field enhancement $d_1$ decreases with growing incident field, i.e., $\left(\partial |d_1| / \partial I_\textrm{inc}\right)_{I_\textrm{inc}\rightarrow0} < 0$, hence, $d_1$ decreases with growing (average) NP temperature. This is mostly because the NP temperature grows as $a^2$, see~\cite{Un-Sivan-size-thermal-effect}. In turn, this causes an increase in the imaginary part of the Au permittivity, which leads to a decrease of the quality factor of the plasmonic resonance, hence, to a broadening of the spectral response and a decrease of the field at resonance. This is inline with the results of~\cite{Gurwich-Sivan-CW-nlty-metal_NP} which were obtained under the quasi-static approximation. Then, for larger NPs, the deterioration of the optical response becomes more modest, until it essentially vanishes. Similarly to the temperature dependence on the NP size, this behaviour can be associated with radiative damping which causes a decrease of the linear thermal response (compare to Fig. 5 in~\cite{Un-Sivan-size-thermal-effect}). In terms of the nonlinear thermo-optic response, this gives rise to an optimal size (here, $a \approx 25$ nm) for which the cubic and quintic corrections to the local-field are maximal (in absolute value) such that a non-perturbative description is required; this is, indeed, reminiscent of the thermal response itself, see~\cite{Un-Sivan-size-thermal-effect}.

The resonance shift of the electric dipole mode (an effect which is removed from Fig.~\ref{fig:d1_Efieldsq}) causes the (linear~\cite{Un-Sivan-size-thermal-effect} and) nonlinear response to drop further (if the illumination is resonant or if $B_m' \Delta T$ shifts the resonance away from the illumination wavelength), and to increase if the system is tuned into resonance. These effects are similar for Ag and Au~\footnote{Here, we ignore dynamic depolarization, which manifests itself only for relatively small NP sizes, and ``quickly'' overwhelmed by radiative damping (see discussion in~\cite{Un-Sivan-size-thermal-effect}).}.

\section{Discussion}\label{sec:discussion}
Our numerical simulations and further analysis unfold the complicated multi-parameter, non-intrinsic dependence of the thermo-optic response of metal NPs. In particular, the contribution of NP size, illumination wavelength and optical and thermal properties of the host explain, at least partially, the variations in reported values of nonlinear response (e.g.,~\cite{plasmonic-SAX-ACS_phot,plasmonic-SAX-PRL,plasmonic-SAX-rods-Ag,Cheshnovsky-SAX,Sivan-Chu-high-T-nl-plasmonics,Gurwich-Sivan-CW-nlty-metal_NP,Hashimoto-nanoscale-cooling,Hashimoto_Opt_Sca_Spec_Thermometry,Orrit-Caldarola_T_measure,Thesis_Carattino,Hache-cubic-metal-nlty,Iranians_kappa_nlty}). This, indirectly, also explains the even more severe discrepancies in reported values of nonlinearity in ultrafast studies (see e.g.,~\cite{Boyd-metal-nlty}) where various additional parameters play a role (most notably, the different electron and phonon temperatures, and the associated heat capacities and $e-ph$ coupling coefficient), and the parametric dependence on the NP size etc. is different. In particular, for small nanoparticles, it was shown in~\cite{Baffou_pulsed_heat_eq_with_Kapitza} that the temperature grows as $a^2$ (sphere area) for CW illumination but as $a^3$ (sphere volume) for ultrafast illumination. All the above shows that it is, in general, {\em incorrect} to deduce any scaling of the steady-state solution from the ultrafast solution.

Our main finding is that the nonlinear steady-state thermo-optical response of metal NPs has the same dependence on the NP size exhibited by the temperature. Indeed, the nonlinearity grows with the NP size, it is highest for NP sizes of a few tens of nms, and then decreases for even larger NPs due to the indirect effect of radiative damping. At the point of strongest response, the numerical examples and consequent approximate analysis above show that the thermo-optical response of metal NPs is remarkably strong, especially considering the associated subwavelength scales involved. In particular, it can reach several {\em hundreds of percent}. This strong nonlinearity contrasts the previous pessimistic claims, made in the context of ultrafast (local) nonlinearities~\cite{Khurgin_chi3} and is a thousand times stronger than that reported previously in strongly nonlinear systems such as $\varepsilon$-near-zero materials~\cite{Boyd_NLO_ENZ_ITO,Boyd_NLO_ENZ_analysis,Shalaev_Faccio_NLO_ENZ}, in which the real part of the permittivity increases by only 0.05\% when the incoming intensity is increased by 1 MW/cm$^2$.

\begin{figure}[h]
\centering
\includegraphics[width=0.5\textwidth]{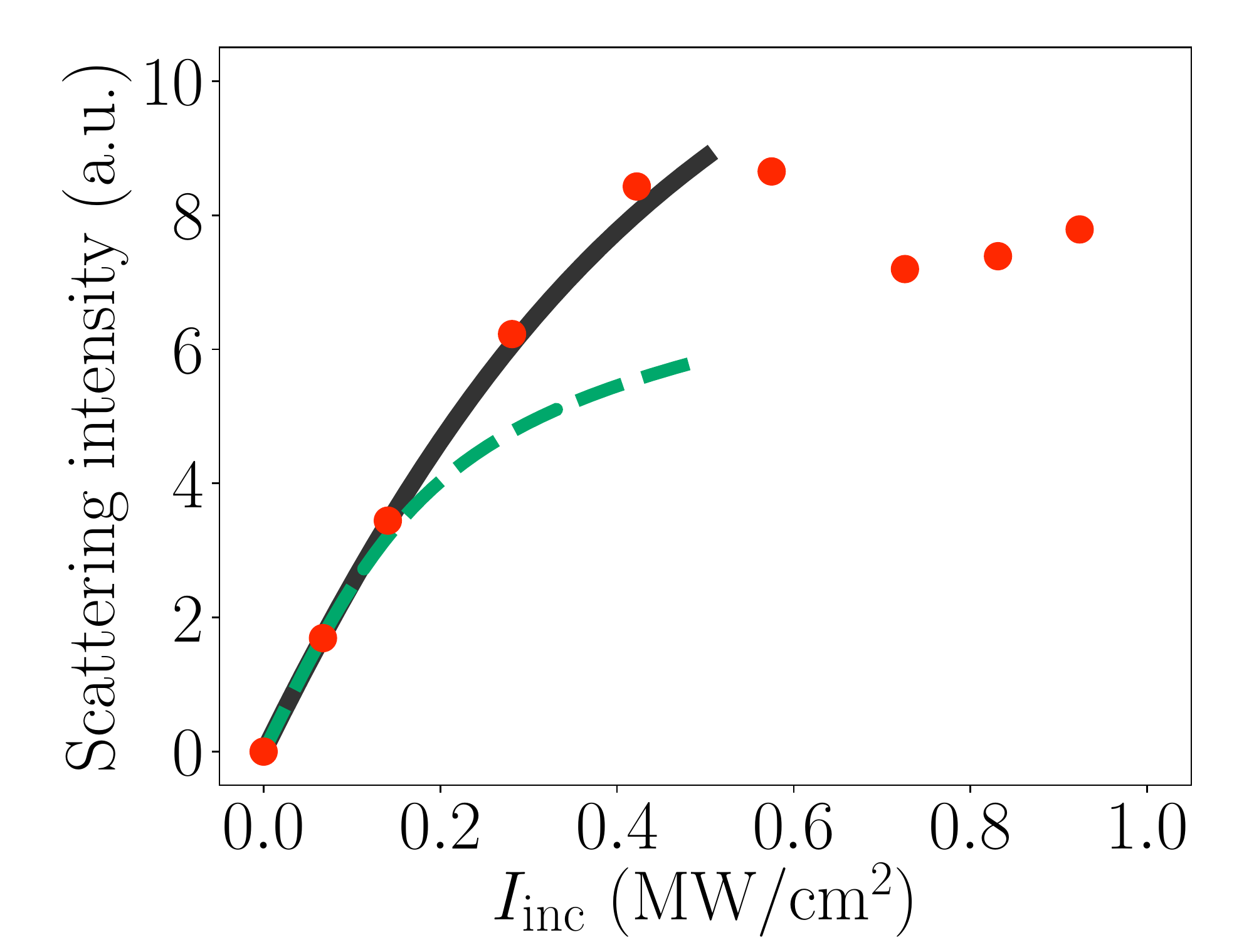}
\caption{(Color online) Nonlinear dependence of the scattering intensity on the incoming intensity for Au NP of 40 nm in radius. The wavelength of the incident light is $560$nm. Red dots represent the experimental data from Ref.~\cite{plasmonic-SAX-ACS_phot}; the black solid line and the green dashed line represent the numerical results based on Eq.~(\ref{eq:tempnpsc}), in which the absorption cross-sections are calculated by the Mie theory and by the quasi-static approximation as in Ref.~\cite{Sivan-Chu-high-T-nl-plasmonics}, respectively. Since the Au NP is immersed in oil and is supported on a glass substrate in the experiments, we use the same configuration as in Section~\ref{sec:config} and Section~\ref{sec:results} in the calculation except that the effective thermal conductivity is set to be $\approx 0.35$ W/(m$\cdot$K)~\cite{Hashimoto-nanoscale-cooling} to mimic the surrounding environment (medium and substrate) in the experiments.}
\label{fig:Isca_sim_vs_exp}
\end{figure}

Potentially the greatest importance of the current study is its ability to improve significantly the agreement between experimental data and modelling of the scattering of intense light from single metal NPs~\cite{plasmonic-SAX-ACS_phot,plasmonic-SAX-PRL,plasmonic-SAX-rods-Ag,Cheshnovsky-SAX,Sivan-Chu-high-T-nl-plasmonics,Gurwich-Sivan-CW-nlty-metal_NP,Jagadale_SWChu_NL_abs_sca_2019,SUSI}; these studies considered initially only backward scattering~\cite{plasmonic-SAX-rods-Ag,plasmonic-SAX-PRL,plasmonic-SAX-OE,plasmonic-SAX-ACS_phot,SUSI}. Indeed, Fig.~\ref{fig:Isca_sim_vs_exp} shows a remarkable quantitative agreement between model and experiment that was absent in previous theoretical studies performed within the limits of quasi-static approximation. This shows that the thermal effect is the most likely source of the optical nonlinearity of metals under CW illumination, thus, resolving the open question raised originally in~\cite{plasmonic-SAX-PRL}; it therefore overall justifies the analysis performed in the current work in its entirety.

Notably, in a later study~\cite{Jagadale_SWChu_NL_abs_sca_2019} the forward scattering was measured as well, such that the dependence of absorption on incoming intensity could be identified experimentally. Specifically, that paper showed that the absorptivity decreases with illumination (when $I_\textrm{inc} > 1.5$ MW/cm$^2$ for NP size of $40$ nm in radius). This effect is not captured by our model, even when extending the modelled temperature regime significantly. Thus, we believe that the reduced absorptivity at high illumination intensities is associated with surface melting that ensues at the edge of the temperature range we study~\cite{Plech_laser_melt_gold_PhysRevB,AuNP_premelt_Nanotech_2008,Nanda_size_dep_mp_NP_2009,ultrafast_laser_melt_AuNP_MD_2011,Kirschner_annurev_physchem,imag_surf_melt_nano_cryst_PNAS_2015,Magnozzi_melt_AuNP_JPCC}. 

In the same vein, we should mention that the current study of the thermo-optic nonlinearity may also not be sufficient to address the so-called ``reverse saturation'' of the scattering/absorption from metal NPs~\cite{plasmonic-SAX-ACS_phot,plasmonic-SAX-PRL,plasmonic-SAX-rods-Ag,Cheshnovsky-SAX}. Specifically, the illumination intensity for the ``reverse saturation'' of the scattering/absorption to occur (e.g. 1.5 - 2 MW/cm$^2$ for Au NPs of $a = 40$ nm) is much higher than the illumination level in the current study. A quantitative agreement thus probably requires taking account of the metal surface melting. For even higher illumination intensities, surface melting leads to NP size reduction~\cite{japanese_size_reduction,Orrit-bubble}.

Additional improvements to the modelling should include accounting for the Kapitza resistance (unless the NPs are relatively large, see~\cite{Sivan-Chu-high-T-nl-plasmonics}), take account of the temperature-dependence of $\varepsilon_h$~\cite{Hashimoto_Opt_Sca_Spec_Thermometry,Stoll_review,Stoll_environment,Del_Fatti_cool_dyn_thm_intf_resis_PRB_2009}, and whenever the NPs are supported on a substrate, an exact knowledge of the contact geometry~\cite{Hashimoto-nanoscale-cooling}; this information is clearly nearly inaccessible experimentally.

Finally, while we focused here on the study of single metal nanospheres under CW illumination, our work is relevant also for nanosecond illumination (and particularly relevant to resolving the controversies associated with bubble formation dynamics~\cite{Baffou-bubble1,Orrit-bubble} and associated sharp spectral features and potential super-resolution~\cite{PA_imaging_Orrit,PT_PA_imaging_Zharov,PT_PA_imaging_Danielli}), for other particle shapes (as e.g., in~\cite{plasmonic-SAX-rods-Ag,Cheshnovsky-SAX}), as well as for other absorbing materials (such as semiconductors~\cite{Lewi_thm_opt_PbTe_meta_atom,Lewi_thm_reconfig_meta_atoms,Shi-Wei-Si-NLTY,Shi-Wei-Si-Nanoscopy} where additional multipole resonances may become important) and multi-particle configurations~\cite{Ndukaife_pht_thm_NP_trap_metasurf_acsnano,Liu_thermal_vs_nonthermal} and the study of their homogenized properties~\cite{Bergman_Levy_CM_general_nlty,Bergman_EMT_general_nlty,Sali_Bergman_EMT_general_nlty,Palpant_EMT_kerr_media}. In the latter cases, the temperature dependence of the host permittivity may become dominant~\cite{Donner_thermal_lensing}.

\acknowledgements The Authors wish to thank J. Baumberg and S.W. Chu for many useful discussions.

\appendix
\section{Detailed analysis of thermo-optical nonlinearities and their size-dependence}\label{app:detail_analys}
In this Appendix, we provide the detailed expression of the  coefficient of $\Delta T_{\textrm{NP}}^2$ in Eq.~(\ref{eq:tempnp_fur_approx}) based on the Pad\'e expansion of $C_{\textrm{abs}}$ given by Eq.~(\ref{eq:cabs_pd}). This coefficient can be expressed in terms of the thermoderivatives of the absorption cross-section (see Eq.~(\ref{eq:tempnp_fur_approx})). By using the chain rule, one can distinguish the various contributions from $B_m^{\prime\ 2}$, $B_m^{\prime\prime\ 2}$, $B_m^{\prime}B_m^{\prime\prime}$, $D_m^{\prime}$ and $D_m^{\prime\prime}$, namely, 
\begin{align}\label{eq:tempnp_fur_approx_3rd_coeff_chain_rule}
&\left[\dfrac{1}{C_{\textrm{abs}}}\dfrac{d^2 C_{\textrm{abs}}}{dT^2} - 2\left( \dfrac{1}{C_{\textrm{abs}}}\dfrac{d C_{\textrm{abs}}}{dT} \right)^2\right]_{\substack{q_1 = 0, \\ T_\textrm{NP}=T_{h,0}}} \\
=& \left[\dfrac{1}{C_{\textrm{abs}}}\dfrac{\partial^2 C_{\textrm{abs}}}{\partial \varepsilon_m^{\prime 2}} - 2\left(\dfrac{1}{C_{\textrm{abs}}}\dfrac{\partial C_{\textrm{abs}}}{\partial \varepsilon_m^{\prime}}\right)^2\right]_{\substack{q_1 = 0,\\T_\textrm{NP}=T_{h,0}}}B_m^{\prime 2}\nonumber\\+&\left[\dfrac{1}{C_{\textrm{abs}}}\dfrac{\partial^2 C_{\textrm{abs}}}{\partial \varepsilon_m^{\prime\prime 2}} - 2\left(\dfrac{1}{C_{\textrm{abs}}}\dfrac{\partial C_{\textrm{abs}}}{\partial \varepsilon_m^{\prime\prime}}\right)^2\right]_{\substack{q_1 = 0, \\ T_\textrm{NP} = T_{h,0}}}B_m^{\prime\prime 2}\nonumber\\ +& 2\left[\dfrac{1}{C_{\textrm{abs}}}\dfrac{\partial^2 C_{\textrm{abs}}}{\partial \varepsilon_m^{\prime}\partial \varepsilon_m^{\prime\prime}} - 2\left(\dfrac{1}{C_{\textrm{abs}}}\dfrac{\partial C_{\textrm{abs}}}{\partial \varepsilon_m^{\prime}}\right)\left(\dfrac{1}{C_{\textrm{abs}}}\dfrac{\partial C_{\textrm{abs}}}{\partial \varepsilon_m^{\prime\prime}}\right)\right]_{\substack{q_1 = 0,\\T_\textrm{NP}=T_{h,0}}}B_m^{\prime}B_m^{\prime\prime} \nonumber \\ + & 2 \left(\dfrac{1}{C_{\textrm{abs}}}\dfrac{\partial C_{\textrm{abs}}}{\partial \varepsilon_m^{\prime}}\right)_{\substack{q_1 = 0,\\T_\textrm{NP}=T_{h,0}}} D_m^{\prime}+2\left(\dfrac{1}{C_{\textrm{abs}}}\dfrac{\partial C_{\textrm{abs}}}{\partial \varepsilon_m^{\prime\prime}}\right)_{\substack{q_1 = 0,\\T_\textrm{NP}=T_{h,0}}}D_m^{\prime\prime}, \nonumber
\end{align}
where the contributions from $B_m^{\prime\ 2}$, $B_m^{\prime\prime\ 2}$ and $B_m^{\prime} B_m^{\prime\prime}$ are respectively given by
\begin{align}\label{eq:tempnp_fur_approx_3rd_coeff_bm1^2}
&\left[\dfrac{1}{C_{\textrm{abs}}}\dfrac{\partial^2 C_{\textrm{abs}}}{\partial \varepsilon_m^{\prime 2}} - 2\left(\dfrac{1}{C_{\textrm{abs}}}\dfrac{\partial C_{\textrm{abs}}}{\partial \varepsilon_m^{\prime}}\right)^2\right]_{\substack{q_1 = 0,\\T_\textrm{NP}=T_{h,0}}}\\ =& -\dfrac{1}{2}\left(\dfrac{1}{C_{\textrm{abs}}}\dfrac{\partial C_{\textrm{abs}}}{\partial \varepsilon_m^{\prime}}\right)_{\substack{q_1 = 0,\\T_\textrm{NP}=T_{h,0}}}^2-\dfrac{1}{2}\left(\dfrac{1}{C_{\textrm{abs}}}\dfrac{\partial C_{\textrm{abs}}}{\partial \varepsilon_m^{\prime\prime}}-\dfrac{1}{\varepsilon_{m}^{\prime\prime}}\right)_{\substack{q_1 = 0,\\T_\textrm{NP}=T_{h,0}}}^2\nonumber\\=&-\dfrac{2}{q_2^2(\varepsilon_{m,0},\varepsilon_h,x)}\left[\left(1-\dfrac{3}{5}x^2\right)^2+\dfrac{4}{9}x^6\right]\nonumber\\
    \rightarrow &\begin{cases}-\dfrac{2}{\varepsilon_{m,0}^{\prime\prime 2}}&\textrm{if }x\rightarrow0\\-\dfrac{2}{(\varepsilon_{m,0}^{\prime}-\varepsilon_h)^2}&\textrm{if } x^3\textrm{ dominates }\end{cases},\nonumber
\end{align}

\begin{align}\label{eq:tempnp_fur_approx_3rd_coeff_bm1_bm2}
&\left[\dfrac{1}{C_{\textrm{abs}}}\dfrac{\partial^2 C_{\textrm{abs}}}{\partial \varepsilon_m^{\prime\prime 2}} - 2\left(\dfrac{1}{C_{\textrm{abs}}}\dfrac{\partial C_{\textrm{abs}}}{\partial \varepsilon_m^{\prime\prime}}\right)^2\right]_{\substack{q_1 = 0,\\T_\textrm{NP}=T_{h,0}}}\\ =& -\dfrac{1}{2}\left(\dfrac{1}{C_{\textrm{abs}}}\dfrac{\partial C_{\textrm{abs}}}{\partial \varepsilon_m^{\prime}}\right)_{\substack{q_1 = 0,\\T_\textrm{NP}=T_{h,0}}}^2-\dfrac{1}{2}\left(\dfrac{1}{C_{\textrm{abs}}}\dfrac{\partial C_{\textrm{abs}}}{\partial \varepsilon_m^{\prime\prime}}+\dfrac{1}{\varepsilon_m^{\prime\prime}}\right)_{\substack{q_1 = 0,\\T_\textrm{NP}=T_{h,0}}}^2\nonumber\\=&-\dfrac{8}{9}\dfrac{x^6}{q_2^2(\varepsilon_{m,0},\varepsilon_h,x)}\left[1+\left(\dfrac{\varepsilon_m^{\prime}-\varepsilon_h}{\varepsilon_m^{\prime\prime}}\right)^2\right]_{\substack{q_1 = 0,\\T_\textrm{NP}=T_{h,0}}}\nonumber\\
    \rightarrow&\begin{cases}0&\textrm{if }x\rightarrow0\\-\dfrac{2}{(\varepsilon_{m,0}^{\prime}-\varepsilon_h)^2}-\dfrac{2}{\varepsilon_{m,0}^{\prime\prime 2}}&\textrm{if } x^3\textrm{ dominates }\end{cases},\nonumber
\end{align}
and 
\begin{align}\label{eq:tempnp_fur_approx_3rd_coeff_bm2^2}
&\left[\dfrac{1}{C_{\textrm{abs}}}\dfrac{\partial^2 C_{\textrm{abs}}}{\partial \varepsilon_m^{\prime}\partial \varepsilon_m^{\prime\prime}} - 2\left(\dfrac{1}{C_{\textrm{abs}}}\dfrac{\partial C_{\textrm{abs}}}{\partial \varepsilon_m^{\prime}}\right)\left(\dfrac{1}{C_{\textrm{abs}}}\dfrac{\partial C_{\textrm{abs}}}{\partial \varepsilon_m^{\prime\prime}}\right)\right]_{\substack{q_1 = 0,\\T_\textrm{NP}=T_{h,0}}}\\ =& -\left[\dfrac{1}{\varepsilon_m^{\prime\prime}}\left(\dfrac{1}{C_{\textrm{abs}}}\dfrac{\partial C_{\textrm{abs}}}{\partial \varepsilon_m^{\prime}}\right)\right]_{\substack{q_1 = 0,\\T_\textrm{NP} = T_{h,0}}} \nonumber \\
=&-\dfrac{4}{3}\dfrac{x^3}{\varepsilon_{m,0}^{\prime\prime} q_2(\varepsilon_{m,0},\varepsilon_h,x)}\nonumber \\
&\rightarrow \begin{cases}0 & \textrm{if }x\rightarrow0\\-\dfrac{2}{\varepsilon_{m,0}^{\prime\prime}|\varepsilon_{m,0}^{\prime}-\varepsilon_h|}&\textrm{if } x^3\textrm{ dominates }\end{cases},\nonumber
\end{align}
$\left(\dfrac{1}{C_\textrm{abs}}\dfrac{\partial C_\textrm{abs}}{\partial \varepsilon_m^{\prime}}\right)_{\substack{q_1 = 0,\\T_\textrm{NP}=T_{h,0}}}$ and $\left(\dfrac{1}{C_\textrm{abs}}\dfrac{\partial C_\textrm{abs}}{\partial \varepsilon_m^{\prime\prime}}\right)_{\substack{q_1 = 0,\\T_\textrm{NP}=T_{h,0}}}$ are respectively given by Eq.~(\ref{eq:cabs_sen_pd_real}) and Eq.~(\ref{eq:cabs_sen_pd_imag}). Therefore, Eq.~(\ref{eq:tempnp_fur_approx_3rd_coeff_chain_rule}) becomes
\begin{align}\label{eq:tempnp_fur_approx_3rd_coeff_x_limit}
&\left[\dfrac{1}{C_{\textrm{abs}}}\dfrac{d^2 C_{\textrm{abs}}}{dT^2} - 2\left( \dfrac{1}{C_{\textrm{abs}}}\dfrac{d C_{\textrm{abs}}}{dT} \right)^2\right]_{\substack{q_1 = 0,\\T_\textrm{NP}=T_{h,0}}}\\ \approx & \begin{cases}-2\left[\left(\dfrac{B_m^{\prime}}{\varepsilon_m^{\prime\prime}}\right)^2+\dfrac{D_m^{\prime\prime}}{\varepsilon_m^{\prime\prime}}\right]& \textrm{if }x\rightarrow0, \\
2\left[-\left(\dfrac{B_m^{\prime}}{|\varepsilon_{m,0}-\varepsilon_h|}+\dfrac{B_m^{\prime\prime}}{\varepsilon_{m,0}^{\prime\prime}}\right)^2 + \dfrac{D_m^{\prime}}{|\varepsilon_m^{\prime}-\varepsilon_h|}+\dfrac{D_m^{\prime\prime}}{\varepsilon_m^{\prime\prime}}\right]& \textrm{if } x^3\textrm{ dominates. }\end{cases}\nonumber
\end{align}

\input{my_bbl.bbl}
\end{document}

%% file: my_bbl.bbl
%